\documentclass[12pt]{article}
\usepackage{"conroy_latex_v2"}
\usepackage{array}

\title{\bf \Large Sensitivity Analysis for Dynamic Discrete Choice Models}
\author{Chun Pong Lau\thanks{\mbox{Kenneth C. Griffin Department of Economics, 
The University of Chicago, \href{mailto:ccplau@uchicago.edu}{ccplau@uchicago.edu}.} \\ I am grateful to Alex Torgovitsky for his continuous guidance and advice. I also thank Michael Dinerstein, Jean-Pierre Dub\'{e}, Jacob Leshno, Ali Horta\c{c}su, Kirill Ponomarev, and Azeem Shaikh, as well as participants of the Econometrics Advising Group, the Econometrics Student Group, the Industrial Organization Working Group, and the Third Year Microeconomics Research Seminar at the University of Chicago for valuable comments. All errors are my own.}}

\usepackage{amssymb}
\usepackage{setspace} 
\usepackage{centernot}
\usepackage{marginnote}
\usepackage{multicol}
\usepackage{lscape}
\usepackage{bbding}

\newcommand{\dt}{d_{\theta}}
\newcommand{\dv}{d_V}
\newcommand{\dg}{d_{\gamma}}
\newcommand{\wt}{\widehat{\theta}(\gamma)}
\newcommand{\wv}{\widehat{V}(\gamma)}
\newcommand{\ts}{\theta^\star}
\newcommand{\vs}{V^\star}
\newcommand{\ls}{\lambda^\star}
\newcommand{\vect}{\mathrm{vec}}

\newcommand{\mc}{\text{MC}}
\newcommand{\rc}{\text{RC}}

\def\minmax{\operatornamewithlimits{min/max}}

\newcommand{\splitatcommas}[1]{%
  \begingroup
  \begingroup\lccode`~=`, \lowercase{\endgroup
    \edef~{\mathchar\the\mathcode`, \penalty0 \noexpand\hspace{0pt plus 1em}}%
  }\mathcode`,="8000 #1%
  \endgroup
}

\usepackage{ragged2e}
\allowdisplaybreaks

\usepackage[sc]{mathpazo}
\begin{document}
\fontfamily{ppl}

\maketitle

\begin{abstract}
In dynamic discrete choice models, some parameters, such as the discount factor, are being fixed instead of being estimated. This paper proposes two sensitivity analysis procedures for dynamic discrete choice models with respect to the fixed parameters. First, I develop a local sensitivity measure that estimates the change in the target parameter for a unit change in the fixed parameter. This measure is fast to compute as it does not require model re-estimation. Second, I propose a global sensitivity analysis procedure that uses model primitives to study the relationship between target parameters and fixed parameters. I show how to apply the sensitivity analysis procedures of this paper through two empirical applications. \par 
\vspace{10pt}
\noindent\textbf{Keywords:} Dynamic discrete choice models, sensitivity analysis, discount factor
\end{abstract}

\newpage

\section{Introduction} \label{sec:intro}
In dynamic discrete choice (DDC) models, some structural parameters are often fixed rather than being estimated. A leading example of such parameters is the discount factor, which is nonparametrically unidentified without further restrictions \citep{rust1994hdbk, magnacthesmar2002ecta}. Apart from DDC models, it is also common to have model parameters being fixed exogenously in calibrating general equilibrium models \citep{dawkinsetal2001hoe}. In the rest of this paper, the parameters that are fixed in the estimation procedure are referred to as fixed parameters.

\par 

The choice of the discount factor in DDC models is usually based on the value used in related papers, some relevant rates of return, or some values larger than 0.9. Researchers may conduct sensitivity analysis by repeating part of the estimation at a few other values of the discount factor. But this can be time-consuming because estimating the full model once can take days or weeks. Hence, the current practice can only offer limited information on how the conclusions are affected by the discount factor due to the high computational cost. \par 

In this paper, I propose new approaches to conduct local and global sensitivity analysis for DDC models with respect to the fixed parameters.

To begin with, I develop a local sensitivity measure that examines the change of the target parameter due to a small change in the fixed parameter. I show that the local sensitivity measure is low-cost to compute because it can be obtained by solving a system of linear equations. Researchers do not need to re-estimate the model in order to compute the local sensitivity measure. In addition, reporting the local sensitivity measure can be more informative than re-estimating the model at a few other values of the fixed parameters because readers can estimate the target parameter at their chosen values of the fixed parameters. I show that the local sensitivity measure can serve as a good local approximation through two empirical applications. \par 

If the literature has some consensus that a certain fixed parameter lies in a tight interval, then local sensitivity analysis can already be informative to approximate how the conclusion changes in such an interval. However, this may not always be the case. For the discount factor, evidence from behavioral economics shows that there can be a lot of variation in the discount factor depending on the context and the sample \citep{fredericketal2002jel}. A recent study by \citet{kongetal2022wp} estimates the discount factor for various consumer goods and reports a wide range of discount factors among different products, from 0.357 (for mayonnaise) to 0.999 (for peanut butter). As a result, researchers may be concerned that their conclusion does not hold at another value of the discount factor. \par

This motivates the global sensitivity analysis in the current paper that examines the structure of DDC models and finds conditions on the model primitives under which the target parameter is monotone in the fixed parameter. Monotonicity can be a useful property because of two computational benefits. First, parameters that are monotone in the fixed parameter are bounded by the endpoints. Second, researchers can easily estimate the breakdown point at which the conclusion changes \citep{horowitzmanski1995ecta, klinesantos2013qe, mastenpoirier2020qe}. In the current practice of sensitivity analysis, researchers typically re-estimate the model at a few (e.g., three) neighboring values of the fixed parameter used for the main analysis because estimating the model once is costly. For instance, \citet{barwickpathak2015rand}, \citet{fowlieetal2016jpe}, and \citet{igami2017jpe} repeat the estimation using discount factors around the one used for the main analysis, and examine how the parameter estimates change with the discount factor. Although monotonic patterns are usually shown, they might not necessarily generalize to the entire support of the discount factor.  Using the discount factor as a leading case of fixed parameters in DDC models, I show that utility can be monotone in the discount factor under some conditions on the transition matrices and conditional choice probabilities. However, counterfactuals are not necessarily monotone in the discount factor, even if utility is monotone in the discount factor. Therefore, I also propose a constrained optimization approach for global sensitivity analysis for more general target parameters and fixed parameters.\par 

As will be discussed in Section \ref{sec:local}, the methodology of this paper is not specific to single-agent DDC models nor the discount factor. The procedures proposed in this paper can be applied to other constrained optimization problems (see problem \eqref{eq:cm} ahead) with a unique solution and fixed parameters. Solving the single-agent DDC model using the full solution method is just an example with such a structure. Some other potential economic applications that contain a constrained optimization structure and a subset of parameters being fixed include: dynamic games \citep{egesdaletal2015qe}, dynamic matching \citep{verdierreeling2021restud, chenchoo2022ej}, international trade \citep{ossa2014aer}, and productivity \citep{yang2021aejma}. See \citet{aguirregabiriaetal2021hdbk} for a recent comprehensive review that contains many DDC examples related to industrial organization.

\subsection{Related literature}

This paper contributes to several strands of literature in economics. First, it is related to the literature on sensitivity analysis. \citet{andrewsetal2017qje} and \citet{honoreetal2020jae} develop measures to analyze the sensitivity of parameter estimates to the moments. More closely related are the papers by  \citet{iskrev2019jedc} and \citet{jorgensen2023restat} that conduct sensitivity analysis with respect to calibrated parameters. The former paper focuses on Bayesian approaches to macroeconomic models. The latter proposes a local sensitivity measure to study the sensitivity with respect to calibrated parameters when the target parameters are estimated from minimizing an unconstrained optimization problem with a Generalized Method of Moments (GMM) objective. The first contribution of the current paper is to provide computationally attractive tools for conducting local sensitivity analysis for estimators obtained from constrained optimization problems. Fixed-point constraints are common in economic problems to represent equilibrium conditions. The current paper focuses on sensitivity analysis with respect to the fixed parameters and is different from some other recent papers in the sensitivity analysis literature such as \citet{armstrongkolesar2021qe} which propose confidence intervals robust to local misspecification for overidentified moment condition models, \citet{bonhommeweidner2022qe} which focus on robustness to misspecification within a larger class of models, and \citet{christensenconnault2022wp} which examine sensitivity to the distribution of the latent variables. \par 

The global sensitivity analysis section of this paper shares a similar theme as the literature on monotone comparative statics (e.g., \citet{topkis1998book}). \citet{light2021mor} is a recent paper that provides conditions for the policy function to be monotone in the discount factor, the parameters in the payoff function, or the transition probability function for Markov decision processes. But \citet{light2021mor} considers a different model than the one in the current paper and does not consider estimation of model parameters. I also do not impose conditions like increasing differences that are required in \citet{light2021mor}.

\subsection{Outline}

The rest of the paper is organized as follows. Section \ref{sec:model} describes the setup and notation. Sections \ref{sec:local} and \ref{sec:global} describe the methodology for local and global sensitivity analysis, respectively. Section \ref{sec:empirical} contains two empirical applications in which I apply the methodology to the seminal bus engine replacement example in \citet{rust1987ecta} and to a recent dynamic matching model in \citet{chenchoo2022ej}. Section \ref{sec:conclusion} concludes. All proofs can be found in the appendix.

\section{Model} \label{sec:model}

In this paper, the \citet{rust1987ecta} model is used as the running example to illustrate the local and global sensitivity analysis procedures. In order to introduce the relevant notations, this section starts by describing the canonical single-agent DDC model and common assumptions. Then, I outline some common solution methods for DDC models and their connection with constrained and unconstrained optimization problems that are relevant for the sensitivity analysis procedures in Sections \ref{sec:local} and \ref{sec:global}.

\subsection{Notations}

Consider a DDC model, where time is indexed by $t = 1, \ldots, T$. In each period $t$, each agent $i \in \cI \equiv \{1, \ldots, N\}$ chooses an action $a_{it} \in \cA \equiv \{0, 1, \ldots, A\}$ to maximize discounted future utility based on the state variables $s_{it}$. The  vector of state variables can be decomposed as $s_{it} \equiv (x_{it}, \epsilon_{it})$, where $x_{it} \in \cX \equiv \{1, \ldots, X\}$ is observable by the agents and the researcher and $\epsilon_{it} \equiv (\epsilon_{0it}, \ldots, \epsilon_{Ait}) \in \bR^{A+1}$ is unobservable to the researcher. In addition, assume that $X$ is finite, $\epsilon_{ait}$ is i.i.d. across agents, choices, and states, and that $\epsilon_{it}$ is continuously distributed and has full support over $\bR^{A+1}$. \par 

Each agent chooses the sequence of actions to maximize discounted future utility:
\[
	\max_{\{a_{it}\}} \bE \left[ \sum^\infty_{t=1} \beta^t \widetilde \pi_i(a_{it}, s_{it};\theta) \right],
\]
where $\beta \in [0, 1)$ is the discount factor common across agents and $\widetilde\pi_i(a_{it}, s_{it}; \theta)$ is the utility function of choosing action $a_{it}$ at state $s_{it}$ and parameterized by $\theta \in \Theta \subseteq \bR^{\dt}$.

\subsection{Single-agent dynamic discrete choice model}
In this section, I consider a single-agent stationary DDC model, with the following standard assumptions (see, e.g., \citet{hortacsujoo2023bk}). I omit the $i$ subscript because there is only one agent.

\begin{assu}[Additive separability]  \label{assu:1}
The utility function can be written as
\[
	\widetilde\pi(a_{t}, s_{t}; \theta) = \pi(a_{t}, x_{t}; \theta) + \epsilon_{t},
\]
where $\pi(a_{t}, x_{t}; \theta)$ is bounded and monotone in $x_{t}$.
\end{assu}

\begin{assu}[i.i.d. error terms]  \label{assu:2} 
For any $\epsilon_{t}$, $\bP[\epsilon_{t+1}|\epsilon_{t}] = \bP[\epsilon_{t+1}]$.
\end{assu}

\begin{assu}[Conditional independence] \label{assu:3}
Given $(x_{t}, a_{t})$ observed, $x_{t+1} \indep (\epsilon_{t}, \epsilon_{t+1})$.
\end{assu}

\par 

Let $\overline{V}(x_{t}, \epsilon_{t})$ be the agent's value function. By Bellman's principle of optimality, the value equation can be written as
\[
	\overline{V}(x_{t}, \epsilon_{t}) = \max_{a \in \cA} \left\{
		\pi(a, x_{t}; \theta) + \epsilon_{at}
		+ \beta \bE[\overline{V}(x_{t+1}, \epsilon_{t+1}) | a , x_{t}]\right\}.
\]
Define the choice-specific value function as
\[
	v(a, x) \equiv 
	\pi(a, x; \theta)
		+ \beta \bE[\overline{V}(x_{t+1}, \epsilon_{t+1}) | a_{t} = a , x_{t} = x],
\]
for any $a \in \cA$ and $x \in \cX$. The conditional choice probability (CCP) of choosing action $a \in \cA$ at state $x \in \cX$ is given by
\[
	\bP[a_{t} = a | x_{t} = x] \equiv 
	\bP\left[\left. a \in \argmax_{j \in \cA} \ \left\{ v(j, x) + \epsilon_{jt} \right\} \right| x_{t} = x \right].
\]

The following assumption on the distributions of the unobservables is standard in the literature.

\begin{assu}[Distribution of the unobservables] \label{assu:4}
$\epsilon_{at}$ follows a mean-zero type-1 extreme value (T1EV) distribution for each $a  \in \cA$ and time $t$.
\end{assu}

Under Assumption \ref{assu:4}, the CCP can be written as
\begin{equation}
	\label{eq:ccp-1}
	\bP[a_{t} = a |  x_{t} = x]
	= \frac{\exp\{ v(a, x)\}}{\sum_{j \in \cA} \exp\{ v(j,x)\}},
\end{equation}
for each $a \in \cA$ and $x \in \cX$. \par

Next, let the state transition be governed by the Markov transition matrix $Q_a$, where the $(x, x')$-entry of $Q_a$ is the probability of transitioning from state $x$ in period $t$ to state $x'$ in period $t+1$ when action $a$ is chosen in period $t$, i.e., $q(x'|x, a) \equiv \bP[x_{t+1} = x' | x_{t} = x, a_{t} = a]$. Let $Q_a(x) \equiv (q(1|x,a), \ldots, q(X|x, a))'$ be the $x$-th row of the transition matrix $Q_a$ for any $a \in \cA$. Define the ex ante value function as $V(x) \equiv \bE[\overline{V}(x, \epsilon_{t})]$, and write $V \equiv (V(1), \ldots, V(X))'$. Then, $V$ satisfies the following fixed-point relationship using Assumption \ref{assu:4} that the unobservables follow a mean-zero T1EV distribution:
\begin{align}
	\label{eq:1-fp-b2}
	V(x) & = \log \left\{ \sum_{a \in \cA}
		\exp \left[  \pi(x, a; \theta) + \beta Q_a(x)'V \right]
	\right\},
\end{align}
for any $x \in \cX$.  Let $\Psi^V$ represents the Bellman operator on the right hand side of \eqref{eq:1-fp-b2}, then the fixed-point relationship \eqref{eq:1-fp-b2} can be summarized as follows
\begin{equation}
	\label{eq:1-fp-b}
	V = \Psi^V(\theta, V; \beta).
\end{equation} 

Finally, this section ends with another representation of the flow utility and its connection with CCP. This representation is useful for global sensitivity analysis in Section \ref{sec:global}. Using Lemma 1 of \citet{arcidiaconomiller2011ecta}, there exists a real-valued function $\psi_a(\cdot)$ such that 
\[
	V(x) = v(x,a) + \psi_a(p(x)),
\]
for any $a \in \cA$ and $x \in \cX$. Let $\pi_a \equiv (\pi(1,a; \theta), \ldots, \pi(X,a; \theta))'$ be the vector of utility functions at action $a \in \cA$. Following the discussion in \citet{kalouptsidietal2021wp, kalouptsidietal2021qe}, the vector $\pi_a$ can be expressed as
\begin{align}
	\label{eq:nonp-util}
	\pi_a = A_a \pi_A + b_a(p),
\end{align}
for any $a \in \cA \backslash \{A\}$, where
\begin{align*}
	A_a & \equiv (I_{X} - \beta Q_a)(I_X - \beta Q_A)^{-1} , \\ 
	b_a(p) & \equiv A_a \psi_A(p) - \psi_a(p).
\end{align*}
Under Assumption \ref{assu:4} that the unobservables follow a mean zero T1EV distribution, it follows that $\psi_a(p) = -\log p_a(x)$ for any $a \in \cA$ and $x \in \cX$ (see also \citet{hotzmiller1993restud}), and $b_a(p)$ can be written as
\[
	b_a(p) = -A_a \log p_A(x) + \log p_a(x),
\]
for any $x \in \cX$ and $a \in \cA \backslash \{A\}$.

\subsection{Solution methods} \label{sec:solnm}
There are different methods to solve DDC models (see \citet{aguirregabiriaetal2021hdbk} and \citet{hortacsujoo2023bk} for details). I briefly outline three common approaches in this section in order to emphasize their structure as constrained or unconstrained optimization problems that would fit into the local sensitivity analysis framework in the next section. \par 

Let $L(\theta, V; \beta)$ be the likelihood function. Here, I introduce $\beta$ as the argument of the likelihood function after the semicolon to indicate that it is a parameter that researchers need to specify in advance, and is fixed throughout the estimation procedure. \par

The nested fixed-point method (NFXP) by \citet{rust1987ecta} involves value function iteration and contains two loops. The inner loop takes the parameter $\theta$ as given and finds the fixed point that solves equation \eqref{eq:1-fp-b}. The outer loop finds the parameter $\theta$ that maximizes the likelihood function. \citet{sujudd2012ecta} show that NFXP is equivalent to solving it by a mathematical program with equilibrium constraints (MPEC):
\begin{align}
	\label{eq:4}
	\begin{split}
		\max_{\theta, V} \quad & L(\theta, V; \beta) \\
		\text{s.t.} \quad & {V} = \Psi^V(\theta, V; \beta).
	\end{split}
\end{align}
Upon convergence, the solution must satisfy the Bellman equations and maximize likelihood. As a result, the system \eqref{eq:4} can be used as a starting point for local sensitivity analysis if the researcher uses NFXP or MPEC to solve the DDC model.

Two-step CCP methods can also be written in a similar manner. Using the \citet{hotzmiller1993restud} inversion, the ex ante value function can be written as
\[
	 V(x) 
	 = v(a, x)  - \log \bP[a_{t} = a|x_{t} = x].
\]
By a suitable normalization, such as $v(A, x) = 0$ for all $x \in \cX$ \citep{hortacsujoo2023bk}, the choice-specific value function can be written as
\begin{equation}
	\label{eq:hm-1}
	v(a, x) = \log \bP[a_{t} = a|x_{t} = x] - \log \bP[a_{t} = A|x_{t} = x],
\end{equation}
for each $a \in \cA \backslash\{A\}$ and $x \in \cX$. Hence, with a given estimator of the CCP, the ex ante value functions can be estimated via \eqref{eq:hm-1} over all choices and states. With the estimated $V$ and substituting the constraint into the objective, it becomes an unconstrained optimization problem. \par

\citet{aguirregabiriamira2002ecta} propose the nested pseudo-likelihood method that iterates on the policy function instead. Their fixed-point equation is written as
\begin{equation}
	\label{eq:npl-1}
	P = \Psi^P(\theta, P; \beta),
\end{equation}
where $\Psi^P$ is the policy function operator. The $K$-stage policy iteration estimator takes the estimator of the policy function from the previous stage $\widehat P^{K-1}$ and solves the following problem that updates the policy function via equation \eqref{eq:npl-1}:
\begin{align*}
	\max_{\theta} \quad & L(\theta, P^K; \beta) \\
	\text{s.t.} \quad & P^K = \Psi^P(\theta, \widehat P^{K-1} ; \beta).
\end{align*}

\subsection{Running example: The bus engine replacement problem} \label{sec:rust-intro}
For the rest of this paper, I provide examples in terms of the seminal bus engine replacement problem in \citet{rust1987ecta}. In this problem, Harold Zurcher, the manager,  observes the bus mileage since the last engine replacement. The bus mileage for each bus $i = 1, \ldots, M$ is denoted by $x_{it}$ and the unobservable state variable is $\epsilon_{it}$. In each period, Zurcher chooses to replace ($a_{it} = 1$) or maintain ($a_{it} = 0$) the bus engine. Assume that the utility functions are the same across the buses. Let $\theta \equiv (\text{MC}, \text{RC})$, $\text{MC}$ be the maintenance cost, $\text{RC}$ be the replacement cost, and $c(x, \text{MC})$ be the cost of maintaining engine at mileage $x = 1, \ldots, X$. The utility function is given by
\[
	\widetilde \pi (a, x, \epsilon_{it} ; \theta) = \pi(a, x; \theta) + \epsilon_{ait},
\]
where $\pi(0, x; \theta) = \text{RC} + c(x, \text{MC})$ and $\pi(1, x; \theta) = c(0, \text{MC})$ for any $x \in \cX$. Here, $Q_0$ and $Q_1$ are the two Markov transition matrices that correspond to the actions that choose to maintain and replace the engine, respectively. Bus mileage is reset to 1 if $a_{it} = 1$. Otherwise, the transition probability of mileage follows a multinomial distribution as below: 
\begin{itemize}
\item If $x \leq X - 2$, the transition probability is
\[
	\bP[x_{it+1} = x' | x_{it} = x, a_{it} = 0]
	= 
	\begin{cases}
		\phi_1 & ,\  x' = x + 1			\\
		\phi_2 & ,\ x' = x + 2			\\
		1 - \phi_1 - \phi_2 &  ,\  x' = x
	\end{cases}.
\]
\item If $x \leq X - 1$, the transition probability is
\[
	\bP[x_{it+1} = x' | x_{it} = x, a_{it} = 0]
	= 
	\begin{cases}
		\phi_1 & ,\  x' = x + 1			\\
		1 - \phi_1 &  ,\  x' = x
	\end{cases}.
\]
\item $\bP[x_{it+1} = X | x_{it} = X, a_{it} = 0] = 1$. 
\end{itemize} 
The fixed-point relationship for the Bellman equation in this example can be written explicitly as follows
\[
	V(x) = 
	\log \left\{
		\exp[\text{RC} + c(x, \text{MC}) + \beta Q_0(x)'V ] +
		\exp[c(0, \text{MC}) + \beta Q_1(x)'V ]
	\right\},
\]
for each $x \in \cX$.

\section{Local sensitivity analysis} \label{sec:local}

Let $V \in \cV \subseteq \bR^{\dv}$ be a vector of auxiliary parameters and $\gamma \in \Gamma \subseteq \bR^{\dg}$ be a vector of fixed parameters. Assume that a researcher is interested in estimating $\theta$ through the following constrained optimization problem by first fixing the parameter $\gamma$ as follows:
\begin{align}
	\label{eq:cm}
	\begin{split}
		\min_{\theta \in \Theta, V \in \cV} \quad & L(\theta, V; \gamma) \\ 
		\text{s.t.} \quad & V = F(\theta, V; \gamma),
	\end{split}
\end{align}
where $L(\theta, V; \gamma)$ is the criterion function, and the constraint $V = F(\theta, V; \gamma) $ describes some fixed-point relationship that captures the equilibrium constraints. \par 

In terms of the DDC model in Section \ref{sec:model}, $\theta$ is the utility parameter, $V$ is the value function, and $\gamma$ is the discount factor. For the NFXP, $L(\theta, V; \gamma)$ corresponds to the likelihood function, and $V = F(\theta, V; \gamma)$ corresponds to the fixed-point equation \eqref{eq:1-fp-b2} based on Bellman optimality. Here, I allow $\gamma$ to be a vector because researchers might fix multiple parameters. For instance, \citet{igami2017jpe} calibrates the discount factor, the rate of change of innovation cost, and the number of potential entrants.\par

Let $(\wt, \wv)$ be the solution obtained from solving the constrained optimization problem \eqref{eq:cm}. Note that the optimal solution has $\gamma$ as an argument because the constrained optimization problem is solved with the pre-specified $\gamma$ that is fixed throughout the estimation procedure. \par

The following assumptions on the constrained optimization problem \eqref{eq:cm} are maintained throughout the paper:
\begin{assu} \label{assu:cm} \text{ }
\begin{enumerate}
	\item \label{assu:cm:1} $L$ and $F$ are continuously differentiable in $\theta$, $V$, and $\gamma$ around $(\wt, \wv)$.
	\item \label{assu:cm:2} $(\wt, \wv)$ is a regular point and is the unique solution to the optimization problem \eqref{eq:cm} and belongs to the interior of $\Theta \times \cV$ for each $\gamma \in \Gamma$.
\end{enumerate}
\end{assu}

Assumption {\color{ucmaroon}\ref{assu:cm}.\ref{assu:cm:1}} ensures that the derivatives in the local sensitivity measure exist. See \citet{rust1988siam} and \citet{norets2010qe} for results on differentiability results related to DDC models. Assumption {\color{ucmaroon}\ref{assu:cm}.\ref{assu:cm:2}} ensures that the first-order condition holds as $(\wt, \wv)$ is a regular point \citep[Chapter 3]{bertsekas1999bk} and that there is a unique solution to the optimization problem regardless of the value of the fixed parameter $\gamma \in \Gamma$.

\subsection{Sensitivity measure}  \label{sec:3.1}
 The gradient of $\wt$ with respect to $\gamma$, i.e.,
\begin{align}
	\label{eq:sen-gradient}
	\frac{\partial \wt}{\partial \gamma'},
\end{align}
can be used as a measure of sensitivity. The $(i, j)$-component of the above matrix measures the change in the $i$-th target parameter for a unit change in the $j$-th fixed parameter.
Depending on the parameter and the context, the following sensitivity measures may be easier to interpret:
\begin{enumerate}
\item The elasticity gives the percentage change in $\wt_i$ for one percentage change in $\gamma_j$. It is defined by
\begin{equation}
	\label{eq:sen-elasticity}
	\frac{\partial \wt_i}{\partial \gamma_j} \frac{\gamma_j}{\wt_i},
\end{equation}
when $\wt_i, \gamma_j \neq 0$.
\item The semi-elasticity gives a unit change in $\wt_i$ for a percentage change in $\gamma_j$. It is defined by
	\begin{equation}
		\label{eq:sen-elasticity2}
		\frac{\partial \wt_i}{\partial \gamma_j} \gamma_j,
	\end{equation}
when $\gamma_j \neq 0$.
\end{enumerate}
The sensitivity measure for $\wv$ can be defined analogously.  \par

I show that computing the local sensitivity measures defined above amounts to solving a linear system of equations. The coefficients and constants in the linear system are evaluated at $(\wt, \wv)$ at the original $\gamma$. Hence, there is no need to re-estimate the model at another value of $\gamma$. \par 

The following proposition summarizes the main result of the local sensitivity analysis procedure. For notational simplicity, I write $(\ts, \vs) \equiv (\wt, \wv)$.

\begin{prop}  \label{prop:1} 
Let Assumption \ref{assu:cm} hold. Denote $\ls$ as the Lagrange multiplier for the constrained optimization problem \eqref{eq:cm}, evaluated at the optimal solution.  The local sensitivity measure of $({\ts}', {\vs}', {\ls}')'$ with respect to $\gamma$, i.e.,  $\frac{\partial \theta}{\partial \gamma'}$, $\frac{\partial V}{\partial \gamma'}$, and $\frac{\partial \lambda}{\partial \gamma'}$, can be obtained by solving the following system of $(2\dv + \dt)$ equations in $(2\dv + \dt)$ unknowns:
\begin{align}
\label{eq:main-sys}
	\begin{array}{rccrccrccl}
		A_{\theta, \theta} 
		& \frac{\partial \theta}{\partial \gamma'} 
		& + 
		& A_{\theta, V}  
		& \frac{\partial V}{\partial \gamma'} 
		& - 
		& (F_\theta)'
		& \frac{\partial \lambda}{\partial \gamma'}
		& = 
		& - A_{\theta, \gamma} \\
		A_{V, \theta} 
		& \frac{\partial \theta}{\partial \gamma'} 
		& + 
		& A_{V, V}  
		& \frac{\partial V}{\partial \gamma'} 
		& +
		& (I_{\dv} - F_V)'
		& \frac{\partial \lambda}{\partial \gamma'}
		& = 
		& - A_{V, \gamma} \\
		F_\theta
		& \frac{\partial \theta}{\partial \gamma'} 
		& + 
		& (F_V - I_{\dv})
		& \frac{\partial V}{\partial \gamma'} 
		& 
		& 
		& 
		& = 
		& - F_{\gamma}, \\
	\end{array}
\end{align}
if the system \eqref{eq:main-sys} has a unique solution, where 
\begin{itemize}
	\item $A_{x,y} \equiv \frac{\partial^2 L}{\partial x \partial y'} - R_{d_x} \frac{\partial \vect[(\frac{\partial F}{\partial x'})']}{\partial y'}$ and $F_x \equiv \frac{\partial F}{\partial x'}$ for $x \in \bR^{d_x}$ and $y \in \bR^{d_y}$.
	\item $\vect(B)$ stacks the columns of the matrix $B \in \bR^{m\times n}$ into a column vector, i.e., $\splitatcommas{\vect(B) \equiv (b_{1,1}, \ldots, b_{1,m}, b_{2,1}, \ldots, b_{2,m}, \ldots, b_{n,1}, \ldots, b_{n,m})'}$.
	\item $R_d \equiv (\lambda_1^\star I_d, \lambda_2^\star I_d, \ldots, \lambda_{\dv}^\star I_d) = {\lambda^\star}' \otimes I_d$, where $\otimes$ denotes Kronecker product, and $I_d$ is the $d\times d$ identity matrix.
	\item All the terms above are evaluated at the optimal solution, e.g., $\frac{\partial^2L}{\partial \theta\partial \theta'} \equiv \left. \frac{\partial^2L}{\partial \theta\partial \theta'} \right|_{(\theta, V) = (\theta^\star, V^\star)}$. 
\end{itemize}
\end{prop}

The proof of Proposition \ref{prop:1} can be found in the appendix. Note that the quantities required to compute the sensitivity measures are either already computed in the model estimation procedure or are fast to compute. Other quantities that are not immediately available can be obtained analytically or numerically without model re-estimation. If one wishes to compute the coefficient terms by numerical derivatives, model re-estimation is not required because the derivatives are evaluated around the optimal solution for the fixed value of $\gamma$. 

As already mentioned in the introduction, Proposition \ref{prop:1} can be applied to other economic problems with a constrained optimization structure and unique optimum. I show in Section \ref{sec:uncons} that Proposition \ref{prop:1} nests local sensitivity analysis for unconstrained optimization problems. Thus, the framework in this section is not specific to the running example of DDC models, the T1EV assumption, or the discount factor. \par

Comparing and reporting the local sensitivity measures have two benefits. First, it can be used to compare the sensitivity of the results with respect to the fixed parameters. Researchers can use this to find the fixed parameters that affect the results the most or determine which of the main results are more sensitive to the fixed parameters. Thus, this can also serve as a guide for more extensive sensitivity analysis. \par 

Second, it can be used as a local approximation of the target parameter at another value of the fixed parameter. The empirical applications in Section \ref{sec:empirical} conduct sensitivity analysis and examine the performance of local approximation through two empirical applications.

The following example demonstrates how the quantities in Proposition \ref{prop:1} can be computed in the context of the \citet{rust1987ecta} model.

\begin{eg} \label{eg:rust-terms} In this example, I show the analytical expressions for the terms in the \citet{rust1987ecta} model that are relevant for the linear system in Proposition \ref{prop:1}. The following derivatives have to be evaluated at the optimal solution with the pre-specified discount factor. I follow the notations introduced in Section \ref{sec:rust-intro} and assume the cost function is given by $c(x, \text{MC}) = -\text{MC}x$. In addition, let $x, y, z \in \cX$ and denote $p(x) \equiv \bP[a_{it} = 1 | x_{it} = x]$ for all $x \in \cX$. The likelihood function can be written as
\[
	L = \sum^M_{i=1} \sum^T_{t=1} \{ a_{it} \log p(x_{it}) + (1 - a_{it}) \log [1 - p(x_{it})] \}.
\]

The relevant second derivatives for the likelihood function are as follows:
\begin{align*}
	\frac{\partial^2 L}{\partial \theta \partial \theta'}
	& = -\sum^M_{i=1} \sum^T_{t=1} p(x_{it})[1 - p(x_{it})] \begin{pmatrix} x_{it}^2 & -x_{it} \\ -x_{it} & 1 \end{pmatrix}, \\
	\frac{\partial^2 L}{\partial \theta \partial V(y)}
	& = -\sum^M_{i=1} \sum^T_{t=1} \beta p(x_{it})[1 - p(x_{it})] [q(y|x_{it}, 0) - q(y|x_{it}, 1)] \begin{pmatrix} -x_{it} \\ 1 \end{pmatrix}, \\ 
	\frac{\partial^2 L}{\partial \theta \partial \beta}
	& = -\sum^M_{i=1} \sum^T_{t=1} \beta p(x_{it})[1 - p(x_{it})] [Q_0(x_{it}) - Q_1(x_{it})]'V  \begin{pmatrix} -x_{it} \\ 1 \end{pmatrix}, \\ 
	\frac{\partial^2 L}{\partial V(y) \partial V(z)}
	& = -\sum^M_{i=1} \sum^T_{t=1} \beta^2 p(x_{it}) [1-p(x_{it})]  [q(y|x_{it}, 0) - q(y|x_{it}, 1)] [q(z|x_{it}, 0) - q(z|x_{it}, 1)] , \\ 
	\frac{\partial^2 L}{\partial V(y) \partial \beta}
	& = -\sum^M_{i=1}\sum^T_{t=1} [q(y|x_{it}, 0) - q(y|x_{it}, 1)] \left\{ 
		[a_{it} - p(x_{it})] + \beta^2 p(x_{it})[1-p(x_{it})] \right\}.
\end{align*}
The relevant first derivatives for the fixed-point equation are as follows:
\begin{align*}
	\frac{\partial F(x)}{\partial \theta}
	& = [1 - p(x)] \begin{pmatrix} -x \\ 1 \end{pmatrix}, \\
	\frac{\partial F(x)}{\partial V(y)}
	& = \beta \{ q(y|x, 0)[1 - p(x)] + q(y|x, 1) p(x) \},\\
	\frac{\partial F(x)}{\partial \beta}
	& = Q_0(x)'V [1 - p(x)] + Q_1(x)'V p(x).
\end{align*}
The relevant second derivatives for the fixed-point equation are as follows:
\begin{align*}
	\frac{\partial^2 F(x)}{\partial \theta \partial \theta'}
	& = p(x)[1 - p(x)] \begin{pmatrix} x^2 & -x \\ -x & 1 \end{pmatrix}, \\
	\frac{\partial^2 F(x)}{\partial \theta \partial V(y)}
	& = \beta [q(y|x, 0) - q(y|x, 1)] p(x)[1 - p(x)] \begin{pmatrix} -x \\ 1 \end{pmatrix}, \\
	\frac{\partial^2 F(x)}{\partial \theta \partial \beta}
	& = p(x)[1 - p(x)] [Q_0(x) - Q_1(x)]'V \begin{pmatrix} -x \\ 1 \end{pmatrix}, \\
	\frac{\partial^2 F(x)}{\partial V(y) \partial V(z)}
	& = \beta^2 [q(y|x, 0) - q(y|x, 1)][q(z|x, 0) - q(z|x, 1)] p(x) [1 - p(x)],\\
	\frac{\partial^2 F(x)}{\partial V(y) \partial \beta}
	& = q(y|x, 0)[1 - p(x)] + q(y|x, 1)p(x) \\
	& \quad + \beta [q(y|x, 0) - q(y|x, 1)] p(x)[1 - p(x)] [Q_0(x) - Q_1(x)]'V.
\end{align*}
\end{eg}

\subsection{Sensitivity of counterfactuals}
In practice, researchers typically first estimate the parameters and then perform counterfactual analysis. Some common types of counterfactuals include changing the utility parameters (e.g., college subsidy program in \citet{keanewolpin1997jpe}), changing the transition probabilities (e.g., demand volatility in \citet{collardwexler2013ecta}), and changing the action and state space (e.g., eliminating patients' actions in \citet{crawfordshum2005ecta}). See \citet{kalouptsidietal2021qe} for a detailed discussion on various types of counterfactuals and some identification results for DDC models. \par 

The local sensitivity of the above counterfactuals can be computed based on the estimates from Proposition \ref{prop:1}. In particular, $\frac{\partial \theta}{\partial \gamma'}$ have already been computed. Then, the sensitivity for the counterfactual parameters $\widetilde \theta$ and the counterfactual value function $\widetilde V$ can be obtained as follows.

\begin{enumerate}
	\item Suppose the counterfactual changes the utility parameters to $\widetilde\theta \equiv H(\theta)$. Then, the sensitivity of the utility parameters under the counterfactual can be computed as
	\begin{equation}
		\label{eq:sen-counterf-1}
		\frac{\partial \widetilde\theta}{\partial \gamma'} = \frac{\partial H}{\partial \theta'} \frac{\partial \theta}{\partial \gamma'}.
	\end{equation}
	The elasticity or semi-elasticity measures similar to the ones introduced at the beginning of Section \ref{sec:3.1} can also be easily computed for the counterfactuals. They can be computed by replacing $\widehat\theta$ with $\widetilde\theta$ in  \eqref{eq:sen-elasticity} and \eqref{eq:sen-elasticity2}, and using \eqref{eq:sen-counterf-1}.
	\item For other types of counterfactuals where the utility parameters are unchanged, $\frac{\partial \widetilde\theta}{\partial \gamma'} = \frac{\partial \theta}{\partial \gamma'}$ holds automatically.
	\item Let $\widetilde V = \widetilde F(\widetilde \theta, \widetilde V; \gamma)$ be the updated equilibrium constraints under the counterfactuals. Depending on the counterfactual, there may be a different number of equations. Thus, the sensitivity of $\widetilde V$ can be computed by substituting $\frac{\partial \widetilde\theta}{\partial \gamma'}$, and solving the following system of equations:
	\[
		\frac{\partial \widetilde F}{\partial \theta'} \frac{\partial \widetilde\theta}{\partial \gamma'}
		+ \left( \frac{\partial \widetilde F}{\partial V'} - I_{\widetilde d_V}\right)\frac{\partial \widetilde V}{\partial \gamma'}
		= - \frac{\partial \widetilde F}{\partial \gamma'}.
	\]
\end{enumerate}

\subsection{Other economic applications} \label{sec:otherapp}
While this paper focuses on DDC models, many other economic applications also have a constrained optimization structure as in equation \eqref{eq:cm} with some parameters being fixed in the estimation procedure. Some details of the examples mentioned in Section \ref{sec:intro} are as follows. \par 

\citet{ossa2014aer} studies the effect of tariffs on welfare. The main optimization problem maximizes welfare subject to equilibrium constraints, such as budget constraints and market clearing conditions. Unlike the other applications, the objective is linear in the variables. The calibrated parameter is the elasticity of substitution of industry varieties $\sigma_s$. Sensitivity analysis with respect to calibrated parameters is conducted by resolving the optimization problem with different values of $\sigma_s$.  \par

\citet{igami2017jpe} studies creative destruction in the hard disk drive industry using a dynamic discrete game model. The model is solved using the nested fixed-point approach. Apart from the discount factor, the rate of change of innovation costs and the number of potential entrants are also fixed in the model. Sensitivity analysis is performed by estimating the full model again at different values of the three fixed parameters.

\citet{yang2021aejma} studies aggregate productivity losses due to misallocation. The main optimization problem maximizes likelihood subject to firm optimality conditions. There are three calibrated parameters. They are the span of control, sectoral capital share, and conversion factor. Again, sensitivity analysis with respect to calibrated parameters is conducted by resolving the optimization problem at different calibrated parameters. 

\citet{chenchoo2022ej} study dynamic matching problems. Similar to dynamic discrete choice problems, the main optimization problem maximizes likelihood subject to fixed-point constraints. There are additional constraints that characterize matching equilibria. The discount factor is fixed at 0.95.

\subsection{Connection with unconstrained optimization problems}		\label{sec:uncons}

In this section, I explain how Proposition \ref{prop:1} is related to unconstrained optimization problems with fixed parameters. This is relevant for solution methods such as \citet{hotzmiller1993restud} and \citet{aguirregabiriamira2002ecta} introduced in Section \ref{sec:solnm}. 

Write $V = \overline F(\theta; \gamma)$, so that $V$ can be expressed as a known function of $\theta$ and $\gamma$. As a result, substituting $V$ into the criterion function yields
\[
	L(\theta, V; \gamma) = L(\theta, \overline F(\theta; \gamma); \gamma) \equiv \overline L(\theta; \gamma).
\]
Hence, optimization problem \eqref{eq:cm} becomes
\begin{align}
	\label{eq:unconst-1}
	\min_{\theta \in \Theta} \quad \overline L(\theta; \gamma). 
\end{align}
Here, the parameter of interest in problem \eqref{eq:unconst-1} is $\theta$ and the relevant local sensitivity measure is $\frac{\partial \theta}{\partial \gamma'}$. The following proposition shows how the local sensitivity measure can be computed and the connection with Proposition \ref{prop:1}. \par

\begin{prop} \label{prop:1b}
Consider the optimization problem \eqref{eq:unconst-1}. Assume that 
\begin{itemize}
	\item $\overline L$ is continuously differentiable in $\theta$ and $\gamma$ around $\wt$.
	\item $\wt$ is the unique solution to the optimization problem \eqref{eq:unconst-1} and belongs to the interior of $\Theta$ for each $\gamma \in \Gamma$.
\end{itemize}
Then, the following statements hold:
\begin{enumerate}
	\item \label{prop:1b1} If $\frac{\partial^2 \overline L}{\partial \theta \partial \theta'}$ is invertible, the local sensitivity measure $\frac{\partial \theta}{\partial \gamma'}$ can be obtained by solving 
\begin{equation}
	\label{eq:vsj-6}
	\frac{\partial^2 \overline L}{\partial \theta \partial \theta'} \frac{\partial \theta}{\partial \gamma'} = -\frac{\partial^2 \overline L}{\partial \theta \partial \gamma'}.
\end{equation} 
	\item \label{prop:1b2} In terms of the notations in Proposition \ref{prop:1},
	\begin{align*}
		\frac{\partial^2 \overline L}{\partial \theta \partial \theta'}
		& = A_{\theta, \theta'} 
	+ \frac{\partial^2 L}{\partial \theta \partial V'} \frac{\partial F}{\partial \theta'} 
	+ \frac{\partial F}{\partial \theta'}\frac{\partial^2 L}{\partial V\partial \theta'} +
	\frac{\partial F}{\partial \theta'} \frac{\partial^2 L}{\partial V \partial V'} \frac{\partial F}{\partial \theta'} , \\
		\frac{\partial^2 \overline L}{\partial \theta \partial \gamma'}
		& = A_{\theta, \gamma'} 
	+ \frac{\partial^2 L}{\partial \theta \partial V'} \frac{\partial F}{\partial \gamma'} 
	+ \frac{\partial F}{\partial \theta'}\frac{\partial^2 L}{\partial V\partial \gamma'} +
	\frac{\partial F}{\partial \theta'} \frac{\partial^2 L}{\partial V \partial V'} \frac{\partial F}{\partial \gamma'}. 
	\end{align*}
\end{enumerate}
All the terms above are evaluated at the optimal solution, e.g., $\frac{\partial^2\overline L}{\partial \theta\partial \theta'} \equiv \left. \frac{\partial^2 \overline L}{\partial \theta\partial \theta'} \right|_{\theta = \theta^\star}$. 
\end{prop}

The assumptions of Proposition \ref{prop:1b} are similar to Assumption \ref{assu:cm} but for unconstrained optimization problems. Part \ref{prop:1b1} of Proposition \ref{prop:1b} follows immediately from total differentiating the first-order conditions of optimization problem \eqref{eq:unconst-1}. 
Part \ref{prop:1b2} of Proposition \ref{prop:1b} shows that the local sensitivity measure for the unconstrained problem \eqref{eq:unconst-1} can be expressed using the terms in Proposition \ref{prop:1} for constrained optimization problem \eqref{eq:cm}. This shows Proposition \ref{prop:1} nests unconstrained optimization problems as a special case.

\begin{re}
\citet{jorgensen2023restat} focuses on the following unconstrained optimization problem with a GMM objective:
\begin{equation}
	\label{eq:jorgensen-gmm}
	\min_{\theta \in \Theta} \ \ g_n(\theta; \gamma)'W_n g_n(\theta; \gamma),
\end{equation}
where $g_n(\theta;  \gamma)$ is some vector-valued functions and $W_n$ is a symmetric positive definite weighting matrix. The unconstrained optimization problem \eqref{eq:jorgensen-gmm} can be written in terms of optimization problem \eqref{eq:cm} based on the preceding discussion. Let $\overline L(\theta; \gamma)$ be the objective function in equation \eqref{eq:jorgensen-gmm}. Since $\frac{\partial \overline L}{\partial \theta} = \left[\frac{\partial g_n(\theta; \gamma)}{\partial \theta'}\right]' W_n g_n(\theta; \gamma) $, the expressions in equation \eqref{eq:vsj-6} can be written as
\begin{align*}
	\frac{\partial^2 \overline L}{\partial \theta \partial \theta'}
	& = \left[\frac{\partial g_n(\theta; \gamma)}{\partial \theta} \right]' W_n \frac{\partial g_n(\theta; \gamma)}{\partial \theta'} + 
	\left[ g_n(\theta; \gamma)' W_n \otimes I_{\dt}\right] \frac{\partial \vect\{[\frac{\partial g_n(\theta; \gamma)}{\partial \theta} ]' \}}{\partial \theta'}, \\
	\frac{\partial^2 \overline L}{\partial \theta \partial \gamma'}
	& = \left[\frac{\partial g_n(\theta; \gamma)}{\partial \theta} \right]' W_n \frac{\partial g_n(\theta; \gamma)}{\partial \gamma'} + 
	\left[ g_n(\theta; \gamma)' W_n \otimes I_{\dt}\right] \frac{\partial \vect\{[\frac{\partial g_n(\theta; \gamma)}{\partial \theta} ]' \}}{\partial \gamma'},
\end{align*}
and they can be evaluated at the optimal solution of \eqref{eq:jorgensen-gmm} for a given $\gamma$. \par 
The above terms match the expressions in Proposition 1 of \citet{jorgensen2023restat}. Thus, the local sensitivity measure in  Proposition \ref{prop:1} also nests the unconstrained version in  \citet{jorgensen2023restat}. As discussed in Section \ref{sec:otherapp}, it is common for economic models to be estimated by constrained optimization with equilibrium constraints. The objective function is not necessarily a GMM objective function in empirical applications, as in \citet{ossa2014aer}.   \end{re}

\section{Global sensitivity analysis} \label{sec:global}
Section \ref{sec:local} has provided a local sensitivity measure that is easy to compute and can be used as a good local approximation of the target parameter around its optimal value at the given fixed parameter. As mentioned in the introduction, researchers usually conduct sensitivity analysis by re-estimating the model at a few other values of the fixed parameter. Even though monotonic patterns are usually shown, they do not necessarily extend to the entire support of the fixed parameters.

\par 

Since the discount factor is the most commonly fixed parameter in DDC models, I start by studying what conditions on model primitives can imply that the flow utility is monotone in the discount factor in DDC models. Then, I focus on the linear-in-parameter utility specification that is common in practice. If the parameters are monotone in the discount factor, it is sufficient to estimate the model at the endpoints of the discount factor to obtain the bounds on the parameter. For more general models, the target parameter may not necessarily be monotone in the fixed parameter. As a result, I propose a constrained optimization approach to compute the bounds of the target parameter over a range of fixed parameters. In the context of DDC models and discount factors, the range of discount factors may be obtained by exclusion restrictions. See \citet{magnacthesmar2002ecta} and \citet{abbringdaljord2020qe} for exclusion restrictions and \citet{kongetal2022wp} for a recent application that estimates the discount factor for various consumer goods.

\subsection{Monotonicity of utility} \label{sec:monotonicity}
In this section, I start from the representation in equation \eqref{eq:nonp-util} that connects flow utility and CCP. This is motivated by two-step approaches that estimate CCP in the first step and the utility parameters in the second step.\par 

Assume that the transition matrices $Q_a$ and CCP $p_a$ are available to the researchers for all $a \in \cA$ in equation \eqref{eq:nonp-util}. In addition, consider the following normalization assumption that is standard in many applications.

\begin{assu} \label{assu:5}
$\pi_A(x) = 0$ for any $x \in \cX$.
\end{assu}

Under Assumptions \ref{assu:4} and \ref{assu:5}, the utility vector at state $a \in \cA \backslash \{A\}$ is given by
\begin{align}
	\pi_a & = A_a \psi_A(p) - \psi_a(p) \notag \\ 
	& = (I_X - \beta Q_a)(I_X - \beta Q_A)^{-1} (-\log p_A) + \log p_a. \label{eq:nonp-util2}
\end{align}

The following proposition shows the sign of the flow utility with respect to the discount factor. 
\begin{prop} \label{prop:nonp1}
Let Assumptions \ref{assu:1} to \ref{assu:4} and \ref{assu:5} hold. The derivative of $\pi_a$ in equation \eqref{eq:nonp-util2} with respect to $\beta$ is
\[
	 	\frac{\partial \pi_a}{\partial \beta}
	=-[Q_a(I_X - \beta Q_A)^{-1} - (I_X - \beta Q_A)^{-1} Q_a] (I_X - \beta Q_A)^{-1} (-\log  p_A),
\]
for any $a \in \cA \backslash \{A\}$.

\end{prop}

The following corollary shows that normalizing the utility to another value gives a similar result. 

\begin{cor} \label{cor:nonp1a} 
Let Assumptions \ref{assu:1} to \ref{assu:4} hold and normalize the utility for choice $A$ as $\pi_A = \overline \pi_A$. Then,
\[
	\frac{\partial \pi_a}{\partial \beta}
	= -[Q_a(I_X - \beta Q_A)^{-1} - (I_X - \beta Q_A)^{-1}Q_A ] (I_X - \beta Q_A)^{-1} (\overline \pi_A - \log p_A) ,
\]
for any $a \in \cA \backslash \{A\}$.
\end{cor}

The above corollary shows that a normalization that is different from the one in Assumption \ref{assu:5} only replaces $(-\log p_A)$ in Proposition \ref{prop:nonp1} by $(\overline \pi_A - \log p_A)$ in Corollary \ref{cor:nonp1a}. Therefore, the remaining results in this section will maintain Assumption \ref{assu:5}. \par
 
Proposition \ref{prop:nonp1} can be simplified under $\rho$-period finite dependence between actions $a$ and $A$. For examples and discussion on finite dependence, see \citet{arcidiaconomiller2011ecta, arcidiaconomiller2019qe} and references therein.

\begin{prop}  \label{prop:nonp2}
Let Assumptions \ref{assu:1} to \ref{assu:4} and \ref{assu:5} hold. 
Suppose there exists $a \in \cA \backslash \{A\}$ and $\rho \in \bN$ such that $Q_a(x) Q_A^\rho = Q_A(x)Q_A^\rho$ for all $x \in \cX$. Then,
\begin{equation}
	\label{eq:prop-nonp2-1}
	\frac{\partial \pi_a}{\partial \beta}
	= -(Q_a- Q_A)(I_X + \beta Q_A + \cdots + \beta^{\rho - 1} Q_A^{\rho - 1})(I_X - \beta Q_A)^{-1} (-\log  p_A).
\end{equation}

\end{prop}

The following corollary shows that if $\rho = 1$, the RHS of equation \eqref{eq:prop-nonp2-1} does not depend on $\beta$. As a result, the slope of flow utility for each state is constant across $\beta \in [0, 1)$.

\begin{cor} \label{prop:nonp-cor1}
Let Assumptions \ref{assu:1} to \ref{assu:4} and \ref{assu:5} hold. 
Suppose that there is one-period dependence between choices $a \in \cA \backslash \{A\}$ and $A$ for all states $x\in\cX$. Then, 
\[
	\frac{\partial \pi_a}{\partial \beta} = (-Q_a + Q_A)(-\log p_A),
\]
for any $a \in \cA \backslash \{A\}$. As a result, for each $a \in \cA \backslash \{A\}$ and $x \in \cX$, $\pi_a(x)$ is either increasing, decreasing, or constant across in $\beta \in [0, 1)$.
\end{cor}

One-period finite dependence is common in economics. This holds in models where one of the actions is a renewal or terminal action. A renewal action is an action that resets the states, such as the option to replace the engine in \citet{rust1987ecta}'s bus engine model. A terminal action is a choice where the optimization problem is ended with no more future actions, such as \citet{pakesetal2007rand}'s firm exit model. The following corollary summarizes the conditions for renewal actions.

\begin{cor} \label{prop:nonp-cor2}
Let Assumptions \ref{assu:1} to \ref{assu:4} and \ref{assu:5} hold. 
Suppose that $p_A(1) > 0$ and action $A$ is a renewal action so that the corresponding transition matrix $Q_A$ is given by
\[
	Q_A = 
		\begin{pmatrix} 1 & 0 & \cdots & 0 \\
		1 & 0 & \cdots & 0 \\
		\vdots & \vdots & \ddots & \vdots \\
		1 & 0 & \cdots & 0 \\
		\end{pmatrix}.
\]
\begin{enumerate}
	\item If $p_A(1) \leq p_A(x)$ for any $x \in \cX $, then $\pi_a$ is nondecreasing in $\beta$ for any $a \in \cA \backslash \{A\}$.
	\item If $p_A(1) \geq p_A(x)$ for any $x \in \cX $, then $\pi_a$ is nonincreasing in $\beta$ for any $a \in \cA \backslash \{A\}$.
\end{enumerate}
\end{cor}

The condition in Corollary \ref{prop:nonp-cor2} is easy to verify because it only depends on the CCP of action $A$ and the presence of a renewal action. \par 

While Corollaries \ref{prop:nonp-cor1} and \ref{prop:nonp-cor2} provide some convenient conditions under which the utility is monotone in $\beta$ under some special conditions, it is important to note that the utilities can still be monotone in the discount factor whenever the transition probabilities and the CCP are such that the derivative in Proposition \ref{prop:nonp1} is always positive or negative, even if the conditions in Corollaries \ref{prop:nonp-cor1} and \ref{prop:nonp-cor2} do not hold.

Although it is possible to obtain the monotonicity of utility under some additional conditions on the model primitives, characterizing the monotonicity for counterfactuals is more challenging due to the nonlinearity of the model. Let $\widetilde V$ be the vector of counterfactual ex ante value functions. Let $\widetilde p_a$ be the counterfactual CCP, $\widetilde Q_a$ be the counterfactual transition matrix, $\widetilde \pi_a$ be the counterfactual utility at action $a \in \cA$. Assume that $\widetilde \pi_a$ is related to $\pi_a$ through the following affine transformation
\[
	\widetilde \pi_a = H_a \pi_a + g_a,
\]
for some $X \times X$ matrix $H_a$ and length $X$ vector $g_a$ that are independent of $\beta$. If $\widetilde \pi_A = 0$, then $\widetilde \pi_a$ can be written as
\begin{align}
	\label{eq:countf-1}
	H_a \pi_a + g_a 
	= \widetilde \pi_a 
	= (I_X - \beta \widetilde Q_a)(I_X - \beta \widetilde Q_A)^{-1} (-\log \widetilde p_A) +  \log \widetilde p_a,
\end{align}
for any $a \in \cA \backslash \{A\}$ using \eqref{eq:nonp-util2}. Taking the derivative with respect to $\beta$ on both sides yields
\begin{align}
	\label{eq:countf-2}
	\begin{split}
	H_a \frac{\partial \pi_a}{\partial \beta}
	& = -[\widetilde Q_a(I_X - \beta \widetilde Q_A)^{-1} - (I_X - \beta \widetilde Q_A)^{-1} \widetilde Q_a] (I_X - \beta \widetilde Q_A)^{-1} (-\log  \widetilde p_A) \\
	& \qquad + (I_X - \beta \widetilde Q_a)(I_X - \beta \widetilde Q_A)^{-1} \left(-\frac{1}{\widetilde p_A} \odot \frac{\partial \widetilde p_A}{\partial \beta}\right) + \frac{1}{\widetilde p_a} \odot \frac{\partial \widetilde p_a}{\partial \beta},
	\end{split}
\end{align}
for any $a \in \cA \backslash \{A\}$, where $\odot$ denotes Hadamard product (entrywise product).

\begin{eg}  \label{eg:4.6}
Consider the \citet{rust1987ecta} model again, so that $\cA = \{0,1\}$, $A = 1$, and $\widetilde p_1(x) = 1 - \widetilde p_0(x)$ for each $x \in \cX$. Then, equation \eqref{eq:countf-1} can be used to characterize the relationship between counterfactual CCP as follows
\begin{align*}
	H_0 \pi_0 + g_0 = [ I_X + \beta (\widetilde Q_1 - \widetilde Q_0)](-\log \widetilde p_1) + \log  \widetilde p_0.
\end{align*}
Thus, equation \eqref{eq:countf-2} can be written as
\[
	H_0 \frac{\partial \pi_0}{\partial \beta}
	= (\widetilde Q_1 - \widetilde Q_0)(-\log \widetilde p_1) 
	- [I_X + \beta(\widetilde Q_1 - \widetilde Q_0)] \left( \frac{1}{\widetilde p_1} \odot \frac{\partial \widetilde p_1}{\partial \beta} \right)
	- \frac{1}{\widetilde p_0} \odot \frac{\partial \widetilde p_1}{\partial \beta}.
\]
where $\widetilde{p}_0 \equiv 1_X - \widetilde{p}_1$.
\end{eg}

The above shows the relationship between the derivatives of counterfactuals with respect to the discount factor in DDC models is nonlinear and may not be easy to characterize.

\begin{re} 
Although the above results maintain the T1EV assumption on the unobservables due to Assumption \ref{assu:4}, the same analysis can be applied to other distributions of unobservables. This is because the mapping $\psi_a(p)$ in equation \eqref{eq:nonp-util2} depends on the distribution of the unobservables and the CCP $p$. The distribution of unobservables is independent of the discount factor, and the CCP is assumed to be known. Hence, $\psi_a(p)$ does not depend on the discount factor. As a result, the results can still be applied by replacing the vector $(-\log p_A)$ by $\psi_A(p)$ for other distributions of the unobservables.
\end{re}

\subsection{Monotonicity with linear-in-parameters utility} \label{sec:lin-in-param}
In this section, I connect the results in the previous section to the linear-in-parameters specification of the utility function that is common in empirical applications.

\begin{assu}[Linear-in-parameters] \label{assu:lin}
For each $a \in \cA \backslash \{A\}$, the flow utility at state $x \in \cX$ is given by 
\[
	\pi_a(x) = \varpi_a(x)'\theta,
\]
where $\theta \in \bR^{\dt}$ and $\varpi_a(x) \equiv (\varpi_{a,1}(x), \ldots, \varpi_{a,\dt}(x))'$ are the corresponding coefficients on $\theta$.
\end{assu}

Let $\Pi_a$ be the $(X \times \dt)$ matrix that collects the coefficients on $\theta$ over all the states at $a \in \cA \backslash \{A\}$, so the $x$-th row represents $\varpi_a(x)'$ for each $a \in \cA \backslash \{A\}$. Then, the vector of utilities for each $a \in \cA \backslash\{A\}$ can be written as
\begin{equation}
	\label{eq:util-mat}
	\pi_a = \Pi_a \theta,
\end{equation}
based on Assumption \ref{assu:lin}. \par 

Let $\widehat p_a$ be the estimator of the CCP for action $a \in \cA \backslash \{A\}$. Then, $\widehat \pi_a$ can be estimated via equation \eqref{eq:nonp-util2}. Let $\widehat \pi$ and $\Pi$ be the matrices that stack $\widehat \pi_a$ and $\Pi_a$ over all $a \in \cA \backslash \{A\}$. Then, the utility parameters can be obtained by the following minimum distance problem
\begin{equation}
	\label{eq:min-dist-1}
	\min_{\theta \in \Theta} \ \ (\widehat \pi - \Pi \theta)' W (\widehat \pi - \Pi\theta),
\end{equation}
where $W$ is a positive definite weighting matrix independent of the discount factor.  \par

The following proposition shows the sensitivity of utility parameters estimated from optimization problem \eqref{eq:min-dist-1} with respect to the discount factor.
\begin{prop} \label{prop:7} 
Let Assumptions \ref{assu:1} to \ref{assu:4}, \ref{assu:5} and \ref{assu:lin} hold. Let $\widehat \theta$  be the parameter estimated from problem \eqref{eq:min-dist-1} and assume that the matrix $\Pi' W\Pi$ is invertible. Then, the derivative of $\widehat \theta$ with respect to $\beta$ is given by
\[
	\frac{\partial \widehat\theta}{\partial \beta}=
	(\Pi'W\Pi)^{-1}\Pi'W \frac{\partial \widehat \pi}{\partial \beta}.
\]
\end{prop}

Proposition \ref{prop:7} shows that the sensitivity of the utility parameters with respect to $\beta$ depends on $\Pi$ and $W$, in addition to $\frac{\partial \widehat\pi}{\partial \beta}$. Hence, even if $\widehat \pi$ is monotone in $\beta$, the choice of the parameterization and the weighting matrix can also affect whether the utility parameters are monotone in $\beta$. The following corollary shows a special case where the weighting matrix is an identity matrix and when the same parameter does not appear in more than one choice.

\begin{cor}  \label{cor:6a}
Let Assumptions \ref{assu:1} to \ref{assu:4}, \ref{assu:5} and \ref{assu:lin} hold. 
Suppose that 
\begin{enumerate}
	\item $\theta$ is partitioned as $\theta = (\theta_0', \ldots, \theta_{A-1}')'$ with $\theta_a \in \bR^{d_a}$ for each $a \in \cA \backslash \{\cA\}$ and $\dt = \sum_{a\in\cA \backslash \{A\}} d_a$.
	\item $\pi_a = \Pi_a\theta_a$ and $\Pi_a$ has full rank.
	\item $W$ is an identity matrix.
\end{enumerate}
Then, 
\[
	\frac{\partial \widehat \theta_a}{\partial \beta} = (\Pi_a' \Pi_a)^{-1} \Pi_a' \frac{\partial \widehat \pi_a}{\partial \beta},
\]
for each $a \in \cA \backslash \{A\}$.
\end{cor}

The above discussion has focused exclusively on the global sensitivity analysis for the discount factor. In practice, researchers may fix some other parameters in the utility function in the estimation procedure. An example is the rate of change of innovation cost that \citet{igami2017jpe} calibrates in the coefficient for the sunk cost parameter that he estimates. Let $\delta \in \bR$ be another parameter that the researcher calibrates apart from the discount factor. The following proposition shows the sensitivity of the utility parameters with respect to $\delta$ estimated from problem \eqref{eq:min-dist-1} when part of $\Pi$ contains $\delta$.

\begin{prop} \label{prop:8} 
Let Assumptions \ref{assu:1} to \ref{assu:4}, \ref{assu:5} and \ref{assu:lin} hold. 
Suppose that part of the matrix $\Pi$ contains fixed parameter $\delta \in \bR$, the matrix $\Pi'W\Pi$ is invertible, and that $\widehat \theta$ is estimated from problem \eqref{eq:min-dist-1}. Then, the derivative of $\widehat \theta$ with respect to $\delta$ is given by
\[
	\frac{\partial \widehat \theta}{\partial \delta}
	= -(\Pi'W\Pi)^{-1} \left[ \Pi' (W' + W) \frac{\partial \Pi}{\partial\delta} \widehat \theta
	- \left(\frac{\partial \Pi}{\partial \delta}\right)'W\widehat \pi   \right].
\]
\end{prop}

Whether monotonicity can be established for other fixed parameters depends on the exact structure of the parameterization. This motivates the estimation approach for global sensitivity analysis in the next section.

\subsection{Other approaches}

Researchers can still estimate the bounds on the target parameter over a certain interval of the fixed parameter through a constrained optimization problem. Following Assumption \ref{assu:cm} that the solution is in the interior and that the optimal solution is a regular point, the first-order condition is satisfied at the optimum. Let $\tau(\theta, V; \gamma)$ be the target parameter the researcher wishes to estimate. Then, the bounds on the target parameter over a pre-specified range of fixed parameters $\gamma \in \overline \Gamma \subseteq \Gamma$ can be estimated by the following optimization problem that estimates the bounds subject to the first-order conditions
\begin{align}
	\label{eq:global-est}
	\begin{split}
		\minmax_{\theta \in \Theta, V \in \cV, \gamma \in \overline \Gamma, \lambda \in \bR^{\dv}}
		\quad & \tau(\theta, V; \gamma)\\
		\text{s.t.} \quad  & \frac{\partial Q}{\partial \theta} - \left( \frac{\partial F}{\partial \theta'} \right)'\lambda = 0 \\
		& \frac{\partial Q}{\partial V'} + \left( I_{d_V} - \frac{\partial F}{\partial V'} \right)'\lambda = 0 \\
		& V = F(\theta, V; \gamma).
	\end{split}
\end{align}

If the researcher is interested in estimating the bounds on the target parameter that depends on the counterfactuals with $\widetilde \theta = H(\theta)$, then optimization problem \eqref{eq:global-est} can be updated as
\begin{align*}
		\minmax_{\theta \in \Theta, V \in \cV, \widetilde V \in \widetilde\cV, \gamma \in \overline \Gamma, \lambda \in \bR^{\dv}}
		\quad & \tau(\theta, V, \widetilde V; \gamma)\\
		\text{s.t.} \quad  & \frac{\partial Q}{\partial \theta} - \left( \frac{\partial F}{\partial \theta'} \right)'\lambda = 0 \\
		& \frac{\partial Q}{\partial V'} + \left( I_{d_V} - \frac{\partial F}{\partial V'} \right)'\lambda = 0 \\
		& V = F(\theta, V; \gamma) \\ 
		& \widetilde V = \widetilde F(H(\theta), \widetilde V; \widetilde \gamma).
\end{align*}
The computational time by treating $\gamma$ as a variable depends on $\overline{\Gamma}$ and the structure of the problem. The next section explores this using the \citet{rust1987ecta} model.

\begin{eg} \label{eg:rust-global} In terms of the \citet{rust1987ecta} model, suppose the researcher is interested in the bounds on the target parameter $\tau(\theta, V; \beta)$ over $\beta \in [\underline{\beta}, \overline{\beta}] \equiv \cB$. Then, optimization problem \eqref{eq:global-est} can be implemented as follows:
\begin{align*}
	\minmax_{(\text{MC}, \text{RC}) \in \Theta, V \in \cV, \beta \in \cB, \lambda \in \bR^{\dv}}\quad & \tau(\theta, V; \beta)\\
	\text{s.t.} \quad 
	& -\sum^M_{i=1} \sum^T_{t=1} [a_{it} - p(x_{it})] x_{it} - \sum^{\dv}_{j=1} \lambda_j[1 - p(j)](- j )= 0 \\
	& - \sum^M_{i=1} \sum^T_{t=1} [a_{it} - p(x_{it})] - \sum^{\dv}_{j=1} \lambda_j[1 - p(j)] = 0 \\
	& - \beta \sum^M_{i=1} \sum^T_{t=1} [a_{it} - p(x_{it})] [q(y|x_{it},0) - q(y|x_{it},1)]  \\
	& \quad  + \lambda_y - \beta \sum^{\dv}_{j=1}
	\lambda_j \{ q(y|j, 0)[1 - p(j)] + q(y|j, 1) p(j) \} 	\tag*{for all $y \in \cX$} \\
	& p(x) = \frac{1}{1 + \exp\{\text{RC} - \text{MC}x + \beta(Q_0(x)-Q_1(x))' V\}} \tag*{for all $x \in \cX$} \\
	& V(x) = \log \left\{ \exp[\text{RC} - \text{MC}x + \beta Q_0(x)'V] + \exp[\beta Q_1(x)'V] \right\} \tag*{for all $x \in \cX$}
\end{align*}
The second and third equations above correspond to the first-order conditions with respect to the two utility parameters. The fourth equation corresponds to the first-order condition with respect to the $x$-th component of the value function. The fifth equation is the equation of the choice probability \eqref{eq:ccp-1}. The last equation corresponds to the fixed-point equation for the Bellman equation as in \eqref{eq:1-fp-b}. The above problem can be implemented in standard software such as Knitro. Researchers can also easily incorporate multi-start in Knitro to optimize the bounds to search for the global optimum.
\end{eg}

Another avenue for conducting global sensitivity analysis is to conduct breakdown analysis \citep{horowitzmanski1995ecta, klinesantos2013qe, mastenpoirier2020qe}. In terms of the fixed parameters in DDC models, breakdown analysis can be used to find the range of fixed parameters such that a certain conclusion holds. More precisely, suppose that the researcher is interested in knowing whether the target parameter is above a threshold, i.e., $\widehat\tau(\gamma) \geq \tau^\star$. Then, the robust region (RR) is the region such that the conclusion holds and is defined as
\[
	\text{RR} \equiv \{ \gamma \in \Gamma : \widehat\tau(\gamma) \geq \tau^\star \},
\]
and the breakdown frontier is the boundary of the robust region. If one further restricts attention to the discount factor so that $\gamma = \beta$, and that $\widehat \tau(\beta)$ is monotone in $\beta$, the breakdown frontier can be found by the bisection method.

\section{Empirical applications} \label{sec:empirical}
In this section, I perform sensitivity analysis of two empirical applications with respect to the discount factor. The first application is the seminal bus engine replacement model in \citet{rust1987ecta}. The second application is a recent dynamic marriage matching model from \citet{chenchoo2022ej}.

\subsection{Application 1: Rust (1987)} \label{sec:rust}

In this section, I conduct local and global sensitivity analysis for the \citet{rust1987ecta} model using the data from the group 4 buses. I assume that the utility function is the same as in Section \ref{sec:rust-intro}, and I specify the cost function as $c(x, \mc) = -\mc x$.

\subsubsection{Local analysis}

This section conducts local sensitivity analysis of various parameters with respect to the discount factor. All estimates in this section are obtained by the NFXP. The parameters I consider are as follows:
\begin{enumerate}
	\item Replacement cost.
	\item Maintenance cost.
	\item \label{counterf:1} Counterfactual conditional choice probability at state 90.
	\item \label{counterf:2} Change in average welfare as defined by $W \equiv \frac{1}{X} \sum^X_{x=1} [\widetilde V(x) - V(x)]$.
\end{enumerate}
For parameters \ref{counterf:1} and \ref{counterf:2} above, the counterfactual I consider is a reduction in maintenance cost of 10\%. To analyze the sensitivity of the parameters at different values of the discount factor, I estimate each parameter using the following discount factors $\beta$: 0.85, 0.9, 0.95, 0.99, 0.999, and 0.9999.  \citet{rust1987ecta} uses $\beta = 0.9999$ in the main results. Then, I estimate the local sensitivity measure using the elasticity measure as in \eqref{eq:sen-elasticity} so that the sensitivity measure can be comparable across different target parameters. Using the point estimate and the local sensitivity measure, I approximate the parameter at another discount factor and report the percentage approximation error. The results are reported in Table \ref{tab:rust-1}.

\begin{table}[!ht]
	\centering
	\caption{Local sensitivity analysis for \citet{rust1987ecta}.}
	\label{tab:rust-1}
	
\begin{tabular}[t]{lccccc}
\toprule
\multicolumn{3}{c}{ } & \multicolumn{3}{c}{\bf Approx. error (in \%) for $\bm{\beta'=\beta-\Delta\beta}$} \\
\cmidrule(l{3pt}r{3pt}){4-6}
\textbf{$\bm{\beta}$} & \textbf{Estimate} & \textbf{Elasticity} & \textbf{$\bm{\Delta \beta = 0.0001}$} & \textbf{$\bm{\Delta \beta = 0.001}$} & \textbf{$\bm{\Delta \beta = 0.01}$}\\
\midrule
\addlinespace[0.3em]
\multicolumn{6}{l}{\textbf{Replacement cost}}\\
\hspace{1em}0.8500 & $ 7.982$ & $   0.246$ & $-1.93 \times 10^{-5}$ & $-1.68 \times 10^{-4}$ & $     -0.018$\\
\hspace{1em}0.9000 & $ 8.151$ & $   0.540$ & $-1.55 \times 10^{-5}$ & $   -0.001$ & $     -0.049$\\
\hspace{1em}0.9500 & $ 8.572$ & $   1.566$ & $ 8.00 \times 10^{-6}$ & $   -0.002$ & $     -0.193$\\
\hspace{1em}0.9900 & $ 9.640$ & $   4.893$ & $-1.28 \times 10^{-4}$ & $   -0.009$ & $     -0.800$\\
\hspace{1em}0.9990 & $10.147$ & $   6.541$ & $-1.50 \times 10^{-4}$ & $   -0.012$ & $     -1.146$\\
\hspace{1em}0.9999 & $10.208$ & $   6.744$ & $-8.85 \times 10^{-5}$ & $   -0.013$ & $     -1.195$\\
\addlinespace[0.3em]
\multicolumn{6}{l}{\textbf{Maintenance cost}}\\
\hspace{1em}0.8500 & $ 0.012$ & $  -4.788$ & $-5.22 \times 10^{-5}$ & $-5.76 \times 10^{-6}$ & $     -0.004$\\
\hspace{1em}0.9000 & $ 0.009$ & $  -6.973$ & $-2.41 \times 10^{-5}$ & $-1.29 \times 10^{-4}$ & $     -0.010$\\
\hspace{1em}0.9500 & $ 0.005$ & $ -11.597$ & $ 5.08 \times 10^{-5}$ & $   -0.001$ & $     -0.052$\\
\hspace{1em}0.9900 & $ 0.003$ & $ -20.712$ & $-1.42 \times 10^{-4}$ & $   -0.005$ & $     -0.379$\\
\hspace{1em}0.9990 & $ 0.002$ & $ -24.235$ & $-1.11 \times 10^{-4}$ & $   -0.009$ & $     -0.633$\\
\hspace{1em}0.9999 & $ 0.002$ & $ -24.619$ & $-9.66 \times 10^{-5}$ & $   -0.010$ & $     -0.675$\\
\addlinespace[0.3em]
\multicolumn{6}{l}{\textbf{Counterfactual choice probability at state 90}}\\
\hspace{1em}0.8500 & $ 0.097$ & $  -1.151$ & $-1.06 \times 10^{-4}$ & $ 4.51 \times 10^{-4}$ & $      0.040$\\
\hspace{1em}0.9000 & $ 0.089$ & $  -1.856$ & $-2.72 \times 10^{-5}$ & $    0.001$ & $      0.072$\\
\hspace{1em}0.9500 & $ 0.078$ & $  -3.355$ & $ 8.37 \times 10^{-5}$ & $    0.002$ & $      0.142$\\
\hspace{1em}0.9900 & $ 0.064$ & $  -6.175$ & $-7.59 \times 10^{-5}$ & $    0.003$ & $      0.298$\\
\hspace{1em}0.9990 & $ 0.061$ & $  -7.271$ & $ 3.63 \times 10^{-5}$ & $    0.004$ & $      0.339$\\
\hspace{1em}0.9999 & $ 0.060$ & $  -7.382$ & $-3.73 \times 10^{-5}$ & $    0.003$ & $      0.333$\\
\addlinespace[0.3em]
\multicolumn{6}{l}{\textbf{Change in average welfare}}\\
\hspace{1em}0.8500 & $ 0.310$ & $   0.332$ & $-3.78 \times 10^{-5}$ & $-3.91 \times 10^{-4}$ & $     -0.039$\\
\hspace{1em}0.9000 & $ 0.322$ & $   1.127$ & $-3.41 \times 10^{-5}$ & $   -0.002$ & $     -0.156$\\
\hspace{1em}0.9500 & $ 0.373$ & $   5.937$ & $-1.02 \times 10^{-4}$ & $   -0.014$ & $     -1.277$\\
\hspace{1em}0.9900 & $ 0.927$ & $  76.967$ & $ -0.008$ & $   -0.769$ & $    -63.845$\\
\hspace{1em}0.9990 & $ 7.480$ & $ 973.503$ & $ -0.972$ & $  -95.023$ & $  -7693.629$\\
\hspace{1em}0.9999 & $73.086$ & $9973.111$ & $-99.484$ & $-9720.724$ & $-784775.081$\\
\bottomrule
\end{tabular}

\end{table}

The local sensitivity measure in Table \ref{tab:rust-1} shows that not all parameters are equally sensitive to the choice of the discount factor. On the other hand, the magnitude of the elasticity is increasing in the discount factor for all four parameters. The last three columns of Table \ref{tab:rust-1} report the performance when the estimate and the local sensitivity measure obtained at $\beta$ is used for local approximation for the target parameter when the discount factor is changed to $\beta ' = \beta - \Delta\beta$. Except for the change in average welfare when the discount factor is close to 1, local approximation gives a less than 1\% approximation error in most cases. An intuition for the larger error in the change in average welfare is that the value function can be interpreted as an infinite sum of the discounted present payoffs. 

Table \ref{tab:rust-1} suggests that researchers should be cautious with the sensitivity of the parameters when the discount factor is large. The discount factor is typically higher when the time between two consecutive periods is lower, or when agents discount the future less. Nevertheless, unless the discount factor is very close to 1, local approximation provides a good approximation of the model parameters without the need to re-estimate the full model. 

\subsubsection{Global analysis} 
\begin{figure}[!ht]
	\centering
	\caption{Utility parameter estimates against the discount factor.}
	\label{fig:rust-global-1}
	\includegraphics[scale=1]{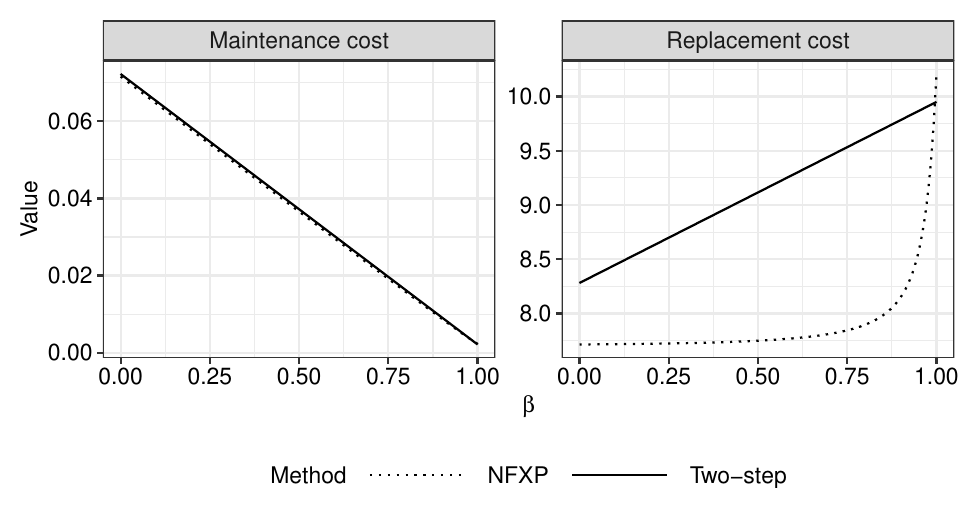}
\end{figure}

In this section, I study the global properties of various target parameters. Figure \ref{fig:rust-global-1} shows the estimates of the maintenance and replacement cost against the discount factor using NFXP and the minimum distance two-step method as in equation \eqref{eq:min-dist-1}. The two-step method in equation \eqref{eq:min-dist-1} requires the CCP to be estimated by data. If the data is rich enough, then the CCP can be estimated by the simple frequency estimator. Since not all 90 states can be observed in the data, I estimate the CCP using a simple logistic regression. I regress the observed action on the state and the square of the state.  \par 
The lines in Figure \ref{fig:rust-global-1} are obtained by estimating the utility parameters for each discount factor in the interval $[0, 0.9999]$ with grid points of size 0.0001. The maintenance cost is strictly decreasing in the discount factor, while the replacement cost is strictly increasing in the discount factor.

\begin{figure}[!ht]
	\centering
	\caption{Estimated probability of engine replacement.}
	\label{fig:rust-global-2}
	\includegraphics[scale=1]{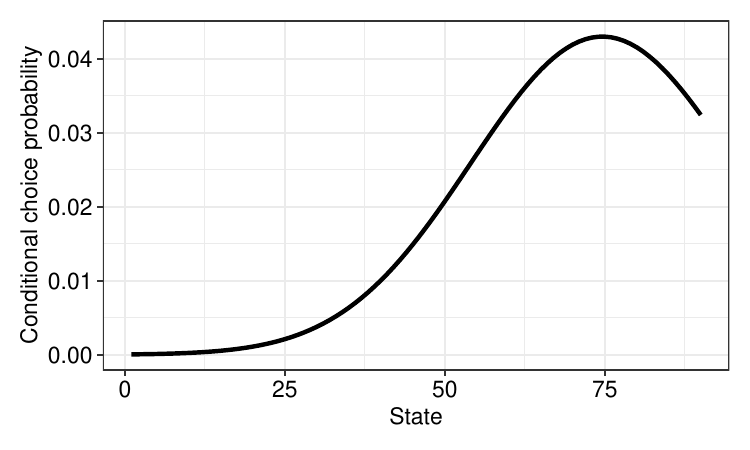}
\end{figure}

Figure \ref{fig:rust-global-2} shows that the estimated CCP satisfies the conditions of Corollary \ref{prop:nonp-cor2}. On the other hand, the option to replace the bus engine is a renewal action. Hence, the flow utility is monotonically increasing in the discount factor.  With the cost function as specified at the beginning of this section, let $\theta = (\text{MC}, \text{RC})$. Then, the matrix of utility coefficients can be written as
\[
	\Pi_0 = 
		\begin{pmatrix} 
			-1 & 1 \\ 
			-2 & 1 \\  
			\vdots & \vdots \\ 
		-90 & 1 \end{pmatrix}.
\]
Applying Corollary \ref{cor:6a}, it follows that $\frac{\partial \widetilde{\text{MC}}}{\partial \beta} < 0$ and $\frac{\partial \widetilde{\text{RC}}}{\partial \beta} > 0$.

\begin{figure}[!ht]
	\centering
	\caption{Counterfactual conditional choice probabilities at states 46, 72, and 90.}
	\label{fig:rust-global-3}
	\includegraphics[scale=1]{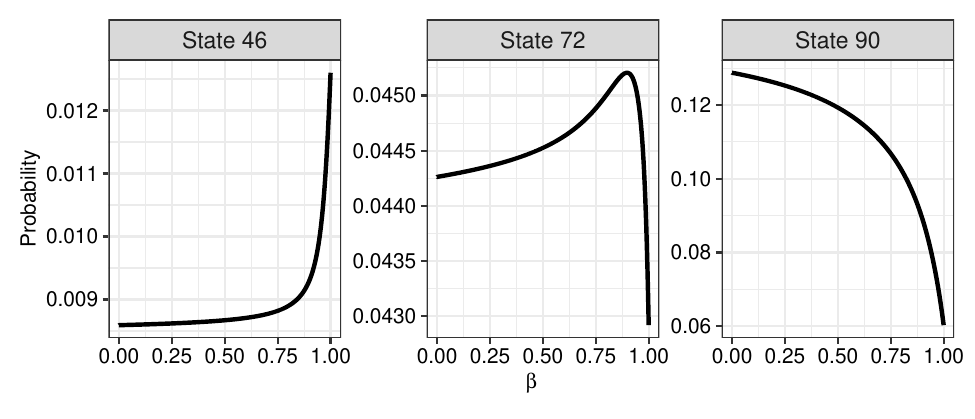}
\end{figure}

Although the utility parameters are monotone in the discount factor, the counterfactuals may not necessarily be monotone in the discount factor. Consider again the counterfactual of reducing the maintenance cost by 10\%. Figure \ref{fig:rust-global-3} shows the counterfactual CCP at three different states. It can be seen that the probabilities can exhibit different properties against the discount factor in Figure \ref{fig:rust-global-3}. Indeed, this can be seen by total differentiating the counterfactual CCP at state $x$, denoted as $\widetilde{p}(x)$, with respect to the discount factor as follows
\begin{equation}
	\label{eq:rust-global-ccp}
	\frac{\partial \widetilde p(x)}{\partial \beta}
	=-{ \underbrace{{\widetilde p(x)} (1-{ \widetilde p(x)})}_{(\text{a})}}
			\Bigg\{
			  { \underbrace{-\frac{\partial \widetilde{\mc}}{\partial \beta} x + \frac{\partial \widetilde{\rc}}{\partial \beta} }_{(\text{b})}}
				+ { \underbrace{[ Q_0(x) - Q_1(x)]' {\left(\widetilde V+ \beta \frac{\partial \widetilde{V}}{\partial \beta} \right)}}_{(\text{c})}}
			 \Bigg\},
\end{equation}
for each $x \in \cX$. In equation \eqref{eq:rust-global-ccp}, (a) is positive. However, (b) and (c) have opposite signs in this model and data. As a result, $\frac{\partial \widetilde p(x)}{\partial \beta}$ is an average of positive and negative terms. Note that $\beta$ also enters the expression in \eqref{eq:rust-global-ccp} explicitly. Hence, the CCP is not necessarily monotone in $\beta$. Figure \ref{fig:rust-global-3} demonstrates the counterfactual CCPs at three different states. Although the counterfactual CCPs demonstrate different shapes, their range is rather small. 

Finally, I explore the performance of using the constrained optimization problem in Example \ref{eg:rust-global} to find the bounds on the target parameter over a range of discount factors. Table \ref{tab:rust-global} shows the bounds on the utility parameters and counterfactuals over different intervals of $\beta$. For all rows, I set the lower bound of the interval of $\beta$ as 0.7 with different upper bounds. The computational time is the total time in running the optimization problem for the lower and upper bounds using Knitro. An attractive feature of the constrained optimization approach is that the optimizer would search for the optimal values so researchers do not need to choose the grid points over the support of the discount factor and re-estimate the model at each of the points. 

\begin{table}[!ht]
	\centering
	\caption{Bounds on two counterfactual target parameters as in Example \ref{eg:rust-global} (the lower bound on $\beta$ is 0.7).}
	\label{tab:rust-global}
	
\begin{tabular}[t]{llc}
\toprule
\textbf{Upper bound on $\bm{\beta}$} & \textbf{Bounds} & \textbf{Computational time (in seconds)}\\
\midrule
\addlinespace[0.3em]
\multicolumn{3}{l}{\textbf{Replacement cost}}\\
\hspace{1em}0.8 & {}[7.810, 7.895] & 23.738\\
\hspace{1em}0.9 & {}[7.810, 8.151] & 19.701\\
\hspace{1em}0.95 & {}[7.810, 8.572] & 24.095\\
\addlinespace[0.3em]
\multicolumn{3}{l}{\textbf{Counterfactual choice probability at state 90}}\\
\hspace{1em}0.8 & {}[0.102, 0.110] & 20.817\\
\hspace{1em}0.9 & {}[0.089, 0.110] & 49.685\\
\hspace{1em}0.95 & {}[0.078, 0.110] & 57.942\\
\bottomrule
\end{tabular}

\end{table}

\subsection{Application 2: Chen and Choo (2023)} \label{sec:cc}

\citet{chenchoo2022ej} develop a dynamic marriage matching framework that extends the methodology in \citet{choo2015ecta}. The empirical example in \citet{chenchoo2022ej} studies how China's one-child policy affects the marriage distribution. The model has a constrained optimization structure, and they estimate their model using the nested fixed-point algorithm. The discount factor is fixed at $\beta = 0.95$ throughout their analysis. I illustrate that the methodology in Section \ref{sec:local} can be applied in this dynamic matching application, although it contains several additional components when compared to the DDC model described in Section \ref{sec:model}. \par

I focus on the analysis in Figure 2 of \citet{chenchoo2022ej}, where they examine how the types of couples affect marital surplus. In their model, individuals differ by education, previous marital status, and age. The education level can be junior high school (JHS), high school (HS), or college (C). Individuals can be previously married and divorced or not married. They report results of ages from 20 to 45 in their Figure 2.\par 

Table 2 of \citet{chenchoo2022ej} compares the surplus differences $\Pi_1(a_{\text{M}}, p_{\text{M}}) - \Pi_0(a_{\text{M}}, p_{\text{M}}, a_{\text{F}}, e_{\text{F}})$. $\Pi_1(a_{\text{M}}, p_{\text{M}})$ is the marriage surplus for an age $a_{\text{M}}$ male with education level JHS, and martial status $p_{\text{M}}$ and an age 20 female with education level JHS, and previously not married. $\Pi_0(a_{\text{M}}, p_{\text{M}}, a_{\text{F}}, e_{\text{F}})$ is the marriage surplus for a male of the same type as in $\Pi_1(a_{\text{M}}, p_{\text{M}})$ and an age $a_{\text{F}}$ female with education level $e_{\text{F}}$ and previously married and divorced. The support of the parameters they consider are $a_{\text{F}} \in \{20, 21, \ldots, 45\}$, $a_{\text{M}} \in \{25, 30, 35\}$, $e_{\text{F}} \in \{\text{JHS}, \text{HS}, \text{C}\}$, and $p_{\text{M}} \in \{0, 1\}$ that equals 1 if the male is previously married and divorced. \citet{chenchoo2022ej} find that couples with similar types generally have a similar surplus.

\begin{figure}[!ht]
	\centering
	\caption{Sensitivity measure for $\Pi_1(a_{\text{M}}, p_{\text{M}}) - \Pi_0(a_{\text{M}}, p_{\text{M}}, a_{\text{F}}, e_{\text{F}})$ estimated at $\beta = 0.95$.}
	\label{fig:cc-2}
	\includegraphics[scale=1]{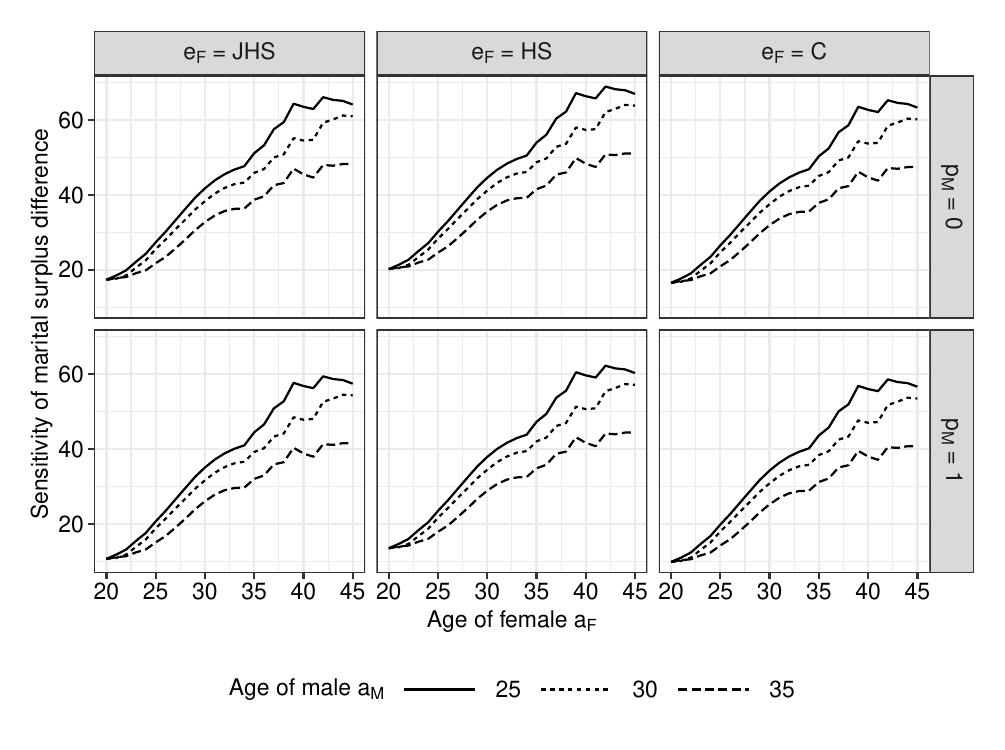}
\end{figure}

Figure \ref{fig:cc-2} shows the sensitivity measure of marital surplus differences as in equation \eqref{eq:sen-gradient} for different types of individuals, arranged in the same format as in Figure 2 of \citet{chenchoo2022ej}. The terms required in Proposition  \ref{prop:1} are computed using automatic differentiation. The sensitivity measure can be interpreted as the unit change in martial surplus for a unit change in the discount factor $\beta$. It can be seen that the sensitivity of the parameter estimates is not constant across different specifications and is generally increasing in $a_{\text{F}}$.

Finally, I examine the performance of the sensitivity measure as a local approximation. To do this, I re-estimate the full model at other values of the discount factor and estimate the surplus difference as in Figure 2 of \citet{chenchoo2022ej}. Then, I approximate the surplus difference at another discount factor using the local sensitivity measure estimated at $\beta = 0.95$. Table \ref{tab:cc-2} reports the summary statistics on the absolute error when local approximation is used to approximate the surplus differences at other discount factors. In the table, $P_x$ refers to the $x$-th percentile. The magnitudes of the absolute errors remain small across different $\beta$ and are mostly less than 1. On the other hand, Table \ref{tab:cc-1} reports the summary statistics of the absolute percentage error for the approximation. It is important to note that the large percentage errors are due to the actual marital surplus differences being very close to 0. See Appendix Figure \ref{fig:cc-supp-1} that shows the large percentage errors are all associated with actual marital surplus differences very close to 0. This exercise shows that the local approximation can be accurate in estimating the surplus without estimating the entire model again at another discount factor.

\begin{table}[!ht]
	\centering
	\caption{Absolute approximation error of approximating marital surplus using the point estimates at $\beta = 0.95$ and the sensitivity measure.}
	\label{tab:cc-2}
	
\begin{tabular}{ccccccccc}
\toprule
\multicolumn{1}{c}{\textbf{ }} & \multicolumn{8}{c}{\textbf{Summary statistics}} \\
\cmidrule(l{3pt}r{3pt}){2-9}
\textbf{$\bm{\beta}$} & \textbf{Min.} & \textbf{$\bm{P_{25}}$} & \textbf{$\bm{P_{50}}$} & \textbf{$\bm{P_{75}}$} & \textbf{$\bm{P_{90}}$} & \textbf{$\bm{P_{95}}$} & \textbf{$\bm{P_{99}}$} & \textbf{Max.}\\
\midrule
0.930 & $0.001$ & $0.049$ & $0.066$ & $0.080$ & $0.090$ & $0.094$ & $0.102$ & $0.107$\\
\cmidrule{1-9}
0.935 & $0.007$ & $0.027$ & $0.035$ & $0.042$ & $0.046$ & $0.050$ & $0.053$ & $0.057$\\
\cmidrule{1-9}
0.940 & $7.36 \times 10^{-5}$ & $0.023$ & $0.037$ & $0.054$ & $0.072$ & $0.110$ & $0.189$ & $0.207$\\
\cmidrule{1-9}
0.945 & $1.13 \times 10^{-5}$ & $0.002$ & $0.004$ & $0.006$ & $0.008$ & $0.009$ & $0.014$ & $0.015$\\
\cmidrule{1-9}
0.955 & $1.32 \times 10^{-4}$ & $0.005$ & $0.013$ & $0.036$ & $0.065$ & $0.089$ & $0.128$ & $0.131$\\
\cmidrule{1-9}
0.960 & $5.96 \times 10^{-5}$ & $0.010$ & $0.021$ & $0.060$ & $0.111$ & $0.149$ & $0.206$ & $0.213$\\
\cmidrule{1-9}
0.965 & $3.72 \times 10^{-4}$ & $0.019$ & $0.032$ & $0.053$ & $0.085$ & $0.121$ & $0.184$ & $0.201$\\
\cmidrule{1-9}
0.970 & $1.65 \times 10^{-4}$ & $0.070$ & $0.157$ & $0.423$ & $0.826$ & $1.043$ & $1.500$ & $1.588$\\
\bottomrule
\end{tabular}

\end{table}

\begin{table}[!ht]
	\centering
	\caption{Absolute approximate percentage error (in \%) of approximating marital surplus using the point estimates at $\beta = 0.95$ and the sensitivity measure.}
	\label{tab:cc-1}
	
\begin{tabular}{ccccccccc}
\toprule
\multicolumn{1}{c}{\textbf{ }} & \multicolumn{8}{c}{\textbf{Summary statistics}} \\
\cmidrule(l{3pt}r{3pt}){2-9}
\textbf{$\bm{\beta}$} & \textbf{Min.} & \textbf{$\bm{P_{25}}$} & \textbf{$\bm{P_{50}}$} & \textbf{$\bm{P_{75}}$} & \textbf{$\bm{P_{90}}$} & \textbf{$\bm{P_{95}}$} & \textbf{$\bm{P_{99}}$} & \textbf{Max.}\\
\midrule
0.930 & $0.004$ & $0.342$ & $0.674$ & $1.500$ & $3.916$ & $ 5.025$ & $20.027$ & $337.273$\\
\cmidrule{1-9}
0.935 & $0.023$ & $0.192$ & $0.330$ & $0.705$ & $1.848$ & $ 2.477$ & $12.738$ & $ 43.120$\\
\cmidrule{1-9}
0.940 & $0.001$ & $0.187$ & $0.420$ & $0.879$ & $1.919$ & $ 3.186$ & $19.946$ & $ 87.474$\\
\cmidrule{1-9}
0.945 & $1.03 \times 10^{-4}$ & $0.017$ & $0.038$ & $0.105$ & $0.274$ & $ 0.389$ & $ 1.573$ & $ 10.163$\\
\cmidrule{1-9}
0.955 & $0.004$ & $0.089$ & $0.189$ & $0.300$ & $0.554$ & $ 0.820$ & $ 1.704$ & $  7.955$\\
\cmidrule{1-9}
0.960 & $0.002$ & $0.124$ & $0.295$ & $0.485$ & $0.821$ & $ 1.333$ & $ 3.737$ & $ 10.974$\\
\cmidrule{1-9}
0.965 & $0.002$ & $0.171$ & $0.335$ & $0.680$ & $1.679$ & $ 3.462$ & $12.618$ & $760.068$\\
\cmidrule{1-9}
0.970 & $0.002$ & $0.955$ & $2.204$ & $3.482$ & $5.982$ & $12.521$ & $49.827$ & $919.764$\\
\bottomrule
\end{tabular}

\end{table}

\section{Conclusion} \label{sec:conclusion}
In this paper, I propose two procedures to conduct sensitivity analysis for parameter estimates with respect to fixed parameters in DDC models. First, the local sensitivity measure reports the change in the target parameter estimated from a constrained optimization problem for a unit change in the fixed parameter. This measure is fast to compute, does not require model re-estimation at another value of the fixed parameter, and nests unconstrained estimation problems. The methodology in this paper can be applied to more general estimation problems, and not necessarily DDC models. For global sensitivity analysis, I examine whether target parameters are monotone in the fixed parameters. Using the discount factor as the leading example of fixed parameters in DDC models, I provide conditions under which utility is monotone in the discount factor.  \par

From the empirical examples, I find that the estimates are typically more sensitive when the discount factor is closer to 1. On the other hand, I also show that the local sensitivity measure can serve as a good local approximation to the target parameter without re-estimating the full model. Hence, researchers can report the local sensitivity measure in addition to the point estimates to increase the transparency of structural research.

\newpage 

\appendix

\numberwithin{equation}{section}
\counterwithin{figure}{section}

\begin{center}
{\bf \Large Appendix}
\end{center}

The appendix contains all the proofs of the main text, and supplemental details on the empirical example on \citet{chenchoo2022ej}.

\section{Proofs} \label{app:A}

\begin{proof}[\textit{\textbf{Proof of Proposition \ref{prop:1}}}]

Consider the optimization problem \eqref{eq:cm}. Let $\lambda$ be the Lagrange multiplier. Then, the Lagrangian can be written as
\begin{equation}
	\cL = L(\theta, V; \gamma) + \lambda' [V - F(\theta, V; \gamma)].
\end{equation}

By Assumption \ref{assu:cm}, the following first-order conditions must hold at the optimum:
\begin{align}
	0 & = \frac{\partial \cL}{\partial \theta'}
	= \frac{\partial L}{\partial \theta} - \left( \frac{\partial F}{\partial \theta'} \right)'\lambda, \label{eq:kkt1} \\
	0 & = \frac{\partial \cL}{\partial V'}
	= \frac{\partial L}{\partial V'} + \left( I_{d_V} - \frac{\partial F}{\partial V'} \right)'\lambda, \label{eq:kkt2} \\
	0 & = \frac{\partial \cL}{\partial \lambda'}
	= V - F. \label{eq:kkt3}  
\end{align}

The following result from \citet[Theorem 9]{magnusneudecker1985jmp} is useful for the remainder of the proof. Let $U \in \bR^{m\times r}$ and $V \in \bR^{r \times p}$. Then, the Jacobian with respect to $X$ can be written as
\begin{equation}
	\label{eq:mn85}
	\frac{\partial \vect(UV)}{\partial [\vect (X)]'}
	= (V' \otimes I_m) \frac{\partial \vect(U)}{\partial [\vect (X)]'} + 
	(I_{p} \otimes U)\frac{\partial \vect(V)}{\partial [\vect (X)]'},
\end{equation}
where $\otimes$ denotes the Kronecker product. \par 
The next step is to total differentiate equations \eqref{eq:kkt1} to \eqref{eq:kkt3} with respect to $\gamma$. Using equation \eqref{eq:mn85} gives
\begin{align*}
	0 
	& = \frac{\partial^2 L}{\partial \theta \partial \theta'} \frac{\partial \theta}{\partial \gamma'} 
	+ \frac{\partial^2 L}{\partial \theta \partial V'} \frac{\partial V}{\partial \gamma'}
	+ \frac{\partial^2 L}{\partial \theta \partial \gamma'} 
	- \left(\frac{\partial F}{\partial \theta'} \right)' \frac{\partial \lambda}{\partial \gamma'} \\
	& \quad -(\lambda' \otimes I_{\dt}) \left\{ \frac{\partial \vect[(\frac{\partial F}{\partial \theta'})']}{\partial \theta'} \frac{\partial \theta}{\partial \gamma'}
	+  \frac{\partial \vect[(\frac{\partial F}{\partial \theta'})']}{\partial V'} \frac{\partial V}{\partial \gamma'}
	+  \frac{\partial \vect[(\frac{\partial F}{\partial \theta'})']}{\partial \gamma'}  \right\}, \\
	0 
	& = \frac{\partial^2 L}{\partial V \partial \theta'} \frac{\partial \theta}{\partial \gamma'} 
	+ \frac{\partial^2 L}{\partial V \partial V'} \frac{\partial V}{\partial \gamma'}
	+ \frac{\partial^2 L}{\partial V \partial \gamma'} 
	 + \left(I_{\dv} - \frac{\partial F}{\partial V'} \right)' \frac{\partial \lambda}{\partial \gamma'} \\
	& \quad -(\lambda' \otimes I_{\dv}) \left\{ \frac{\partial \vect[(\frac{\partial F}{\partial V'})']}{\partial \theta'} \frac{\partial \theta}{\partial \gamma'}
	+  \frac{\partial \vect[(\frac{\partial F}{\partial V'})']}{\partial V'} \frac{\partial V}{\partial \gamma'}
	+  \frac{\partial \vect[(\frac{\partial F}{\partial V'})']}{\partial \gamma'}  \right\}, \\
	0 & = \frac{\partial V}{\partial \gamma'} 
	- \left(\frac{\partial F}{\partial \theta'} \frac{\partial \theta}{\partial \gamma'}
	+ \frac{\partial F}{\partial V'} \frac{\partial V}{\partial \gamma'}
	+ \frac{\partial F}{\partial \gamma'} \right).
\end{align*}
The results follow from rearranging the above equations.
\end{proof}

\begin{proof}[\textit{\textbf{Proof of Proposition \ref{prop:1b}}}]
Under the assumptions of the proposition, the following first-order condition for problem \eqref{eq:unconst-1}  hold at the optimum
\begin{equation}
	\label{eq:prop1b-1}
	\frac{\partial \overline L}{\partial \theta'} = 0.
\end{equation}
Total differentiating equation \eqref{eq:prop1b-1} with respect to $\gamma$ gives
\[
\frac{\partial^2 \overline L}{\partial \theta \partial \theta'} \frac{\partial \theta}{\partial \gamma'} = -\frac{\partial^2 \overline L}{\partial \theta \partial \gamma'}.
\]
This proves part \ref{prop:1b1} of the proposition. \par 

Next, note that $V$ is no longer an argument of $\overline F$. Hence, the linear system in Proposition \ref{prop:1} becomes
\begin{align}
\label{eq:vsj-1}
\arraycolsep=.8pt\def\arraystretch{1.7}
	\begin{array}{rccrccrccl}
		\left\{\frac{\partial^2 L}{\partial \theta\partial \theta'} - R_{\dt}\frac{\partial \vect[(\frac{\partial F}{\partial\theta'})']}{\partial \theta'}\right\}
		& \frac{\partial \theta}{\partial \gamma'} 
		& + 
		& 
		\frac{\partial^2 L}{\partial \theta\partial V'}  
		& \frac{\partial V}{\partial \gamma'} 
		& - 
		& \left(\frac{\partial F}{\partial \theta'}\right)'
		& \frac{\partial \lambda}{\partial \gamma'}
		& = 
		& R_{\dt} \frac{\partial \vect[(\frac{\partial F}{\partial \theta'})']}{\partial \gamma'}  - \frac{\partial^2 L}{\partial \theta \partial \gamma'} \\
		\frac{\partial^2 L}{\partial V\partial \theta'} 
		& \frac{\partial \theta}{\partial \gamma'} 
		& + 
		& 
		\frac{\partial^2 L}{\partial V\partial V'}  
		& \frac{\partial V}{\partial \gamma'} 
		& +
		& 
		& \frac{\partial \lambda}{\partial \gamma'}
		& = 
		& 
		 - \frac{\partial^2 L}{\partial V \partial \gamma'} \\
		\frac{\partial F}{\partial \theta'}
		& \frac{\partial \theta}{\partial \gamma'} 
		& -
		&
		& \frac{\partial V}{\partial \gamma'} 
		& 
		& 
		& 
		& =
		& -\frac{\partial F}{\partial \gamma'}.
	\end{array}
\end{align}
Combining the first two equations in system \eqref{eq:vsj-1}, the variable $\frac{\partial \lambda}{\partial \gamma'}$ can be eliminated. Rearranging the terms after the combination gives
\begin{align}
	\label{eq:vsj-2}
	\begin{split}
		& \left\{
			\frac{\partial^2 L}{\partial \theta\partial \theta'} 
			- R_{\dt}\frac{\partial \vect[(\frac{\partial F}{\partial\theta'})']}{\partial \theta'}
			+  \left(\frac{\partial F}{\partial \theta'}\right)'
			\frac{\partial^2 L}{\partial V\partial \theta'}
		\right\}
		\frac{\partial \theta}{\partial \gamma'} 
		+
		\left[
			\frac{\partial^2 L}{\partial \theta\partial V'} 
			+
			 \left(\frac{\partial F}{\partial \theta'}\right)'
			\frac{\partial^2 L}{\partial V\partial V'}
		\right]
		 \frac{\partial V}{\partial \gamma'}  \\
		 & \quad =
		  R_{\dt} \frac{\partial \vect[(\frac{\partial F}{\partial \theta'})']}{\partial \gamma'}  - \frac{\partial^2 L}{\partial \theta \partial \gamma'} 
		  -  \left(\frac{\partial F}{\partial \theta'}\right)'
		 \frac{\partial^2 L}{\partial V \partial \gamma'}.
	\end{split}
\end{align}
Substituting the last equation in system \eqref{eq:vsj-1} to equation  \eqref{eq:vsj-2}, the $\frac{\partial V}{\partial \gamma'}$ term can be eliminated. This yields
\begin{align}
	\label{eq:vsj-3}
	\begin{split}
		& \left\{
			\frac{\partial^2 L}{\partial \theta\partial \theta'} 
			- R_{\dt}\frac{\partial \vect[(\frac{\partial F}{\partial\theta'})']}{\partial \theta'}
			+  \left(\frac{\partial F}{\partial \theta'}\right)'
			\frac{\partial^2 L}{\partial V\partial \theta'}
			+
				\left[
			\frac{\partial^2 L}{\partial \theta\partial V'} 
			+
			 \left(\frac{\partial F}{\partial \theta'}\right)'
			\frac{\partial^2Q}{\partial V\partial V'}
		\right] \frac{\partial F}{\partial \theta'}
		\right\} 
		\frac{\partial \theta}{\partial \gamma'}  \\
		 & \quad =
		  R_{\dt} \frac{\partial \vect[(\frac{\partial F}{\partial \theta'})']}{\partial \gamma'}  - \frac{\partial^2 L}{\partial \theta \partial \gamma'} 
		  -  \left(\frac{\partial F}{\partial \theta'}\right)'
		 \frac{\partial^2 L}{\partial V \partial \gamma'}
		 - 
		\left[
			\frac{\partial^2 L}{\partial \theta\partial V'} 
			+
			 \left(\frac{\partial F}{\partial \theta'}\right)'
			\frac{\partial^2 L}{\partial V\partial V'}
		\right]  \frac{\partial F}{\partial \gamma'} .
	\end{split}
\end{align}

Using equation \eqref{eq:kkt2}, the Lagrange multiplier can be written as $\lambda = -\frac{\partial L}{\partial V}$, so $-R_{\dt} = (\frac{\partial L}{\partial V})'\otimes I_{\dt}$. In addition, since $F$ is not a funcion of $V$ under the present setup, $\frac{\partial V}{\partial \theta'} = \frac{\partial F}{\partial \theta'}$ holds. It follows that
\begin{align}
	\frac{\partial^2 \overline L}{\partial \theta \partial \theta'}
	& = \frac{\partial}{\partial \theta'} \left(
		\frac{\partial L}{\partial \theta} + \frac{\partial V}{\partial \theta'} \frac{\partial L}{\partial V} 
		\right) \notag \\ 
	& =  \frac{\partial^2 L}{\partial \theta \partial \theta'} 
	+ \frac{\partial^2 L}{\partial \theta \partial V'} \frac{\partial V}{\partial \theta'} 
	+ \left[\left(\frac{\partial L}{\partial V}\right)' \otimes I_{\dt}\right]
	\frac{\partial \vect[(\frac{\partial V}{\partial \theta'})']}{\partial \theta'}
	+ \left(\frac{\partial V}{\partial \theta'}\right)' \left(\frac{\partial^2 L}{\partial V\partial \theta'} +
	\frac{\partial^2 L}{\partial V \partial V'} \frac{\partial V}{\partial \theta'} \right)  \notag \\
	& = \frac{\partial^2 L}{\partial \theta \partial \theta'} 
	+ \frac{\partial^2 L}{\partial \theta \partial V'} \frac{\partial F}{\partial \theta'} 
	- R_{\dt} \frac{\partial \vect[(\frac{\partial V}{\partial \theta'})']}{\partial \theta'}
	+\left( \frac{\partial F}{\partial \theta'} \right)'\frac{\partial^2 L}{\partial V\partial \theta'} +
	\left(\frac{\partial F}{\partial \theta'}\right)' \frac{\partial^2 L}{\partial V \partial V'} \frac{\partial F}{\partial \theta'} , \label{eq:vsj-4}  
\end{align}
and similarly,
\begin{align}
	\frac{\partial^2 \overline L}{\partial \theta \partial \gamma'}
	& = \frac{\partial}{\partial \gamma'} \left(
		\frac{\partial L}{\partial \theta} + \frac{\partial V}{\partial \theta'} \frac{\partial L}{\partial V} 
		\right) \notag \\ 
	& =  \frac{\partial^2 L}{\partial \theta \partial \gamma'} 
	+ \frac{\partial^2 L}{\partial \theta \partial V'} \frac{\partial V}{\partial \gamma'} 
	+ \left[\left(\frac{\partial L}{\partial V}\right)' \otimes I_{\dt}\right]
	\frac{\partial \vect[(\frac{\partial V}{\partial \theta'})']}{\partial \gamma'}
	+ \left(\frac{\partial V}{\partial \theta'} \right)'\left(\frac{\partial^2 L}{\partial V\partial \gamma'} +
	\frac{\partial^2 L}{\partial V \partial V'} \frac{\partial V}{\partial \gamma'} \right)  \notag \\
	& = \frac{\partial^2 L}{\partial \theta \partial \gamma'} 
	+ \frac{\partial^2 L}{\partial \theta \partial V'} \frac{\partial F}{\partial \gamma'} 
	- R_{\dt} \frac{\partial \vect[(\frac{\partial V}{\partial \theta'})']}{\partial \gamma'}
	+ \left(\frac{\partial F}{\partial \theta'} \right)' \frac{\partial^2 L}{\partial V\partial \gamma'} +
	\left(\frac{\partial F}{\partial \theta'} \right)' \frac{\partial^2 L}{\partial V \partial V'} \frac{\partial F}{\partial \gamma'}. \label{eq:vsj-5}  
\end{align}
Substituting equations \eqref{eq:vsj-4}  and  \eqref{eq:vsj-5} into equation  \eqref{eq:vsj-3}  gives  
\begin{equation*}
	\frac{\partial^2 \overline L}{\partial \theta \partial \theta'} \frac{\partial \theta}{\partial \gamma'} = -\frac{\partial^2 \overline L}{\partial \theta \partial \gamma'}.
\end{equation*} 
This proves part \ref{prop:1b2} of the proposition. 
\end{proof}

\begin{proof}[\textbf{\textit{Proof of Proposition \ref{prop:nonp1}}}] To begin with, note that $ p_a$ and $Q_a$ is not a function of $\beta$ for any $a \in \cA$. They are assumed to be known to the researcher. Therefore, the derivative of $\pi_a$ with respect to $\beta$ is 
\begin{equation}
	\label{eq:nonp-1-1}
	\frac{\partial \pi_a}{\partial \beta}
	= \frac{\partial}{\partial \beta}[(I_X - \beta Q_a)(I_X - \beta Q_A)^{-1}]   (-\log p_A),
\end{equation}
for any $a \in \cA \backslash \{A\}$. The derivative can be computed as follows
\begin{align}
		& \hspace{-20pt} \frac{\partial}{\partial \beta}[(I_X - \beta Q_a)(I_X - \beta Q_A)^{-1}]  \notag \\
		& = -Q_a (I_X - \beta Q_A)^{-1} + (-1) (I_X - \beta Q_a)(I_X - \beta Q_A)^{-1} (-Q_A) (I_X - \beta Q_A)^{-1}  \notag \\
		& = -[Q_a - (I_X - \beta Q_a)(I_X - \beta Q_A)^{-1}Q_A ] (I_X - \beta Q_A)^{-1}  \notag \\
		& = -[Q_a - (I_X - \beta Q_A)^{-1}Q_A + \beta Q_a (I_X - \beta Q_A)^{-1} Q_A] (I_X - \beta Q_A)^{-1}   \notag \\
		& = -[Q_a(I_X - \beta Q_A)^{-1} (I_X - \beta Q_A + \beta Q_A) - (I_X - \beta Q_A)^{-1}Q_A ] (I_X - \beta Q_A)^{-1}  \notag \\
		& = -[Q_a(I_X - \beta Q_A)^{-1} - (I_X - \beta Q_A)^{-1}Q_A ] (I_X - \beta Q_A)^{-1}.  \label{eq:deri-quad-term}
	\end{align}
Therefore, the result follows by substituting equation \eqref{eq:deri-quad-term} into equation \eqref{eq:nonp-1-1}.
\end{proof}

\begin{proof}[\textit{\textbf{Proof of Corollary \ref{cor:nonp1a}}}] Under the normalization that $\pi_A = \overline \pi_A$, 
\begin{align}
	\label{eq:cor-nonp1a-1}
	\pi_a & = A_a \overline \pi_A + A_a \psi_A(p) - \psi_a(p) = A_a (\overline \pi_A - \log p_A) + \log p_a,
\end{align}
for any $a \in \cA \backslash \{A\}$ by equation \eqref{eq:nonp-util}. \par 

Note that in equation \eqref{eq:cor-nonp1a-1}, only $A_a$ is a function of $\beta$. Hence, equation \eqref{eq:deri-quad-term} can be applied to show that 
\[
	\frac{\partial \pi_a}{\partial \beta}
	= -[Q_a(I_X - \beta Q_A)^{-1} - (I_X - \beta Q_A)^{-1}Q_A ] (I_X - \beta Q_A)^{-1} (\overline \pi_A - \log p_A) ,
\]
for any $a \in \cA \backslash \{A\}$. 
\end{proof}

\begin{proof}[\textit{\textbf{Proof of Proposition \ref{prop:nonp2}}}]
Let $\{\nu_i\}^X_{i=1}$ be the eigenvalues of $Q_A$. Since $Q_A$ is a Markov matrix, the largest eigenvalue of $Q_A$ is at most 1. Thus, the eigenvalues of $I_X - \beta Q_A$ are $\{1 - \beta \nu_i\}^X_{i=1}$. Since $\beta < 1$, it follows that $1 - \beta \nu_i > 0$ for any $i = 1, \ldots, X$. Hence, $I_X - \beta Q_A$ is invertible. Next, note that 
	\begin{equation}
		\label{eq:nonp-2-1}
		(I_X - \beta Q_A)^{-1} = I_X + \sum^\infty_{r = 1} \beta^r Q_A^r.
	\end{equation}
It follows that the two matrix products can be written as
	\begin{align}
		Q_a (I_X - \beta Q_A)^{-1} 
		= Q_a \left(I_X + \sum^\infty_{r=1} \beta^r Q^r_A\right)
		= Q_a + \sum^\infty_{r=1} \beta^r Q_a Q_A^r,
		\label{eq:mat-inv-1}
	\end{align}
	and
	\begin{equation}
		\label{eq:mat-inv-2}
		 (I_X - \beta Q_A)^{-1} Q_A
		= \left(I_X + \sum^\infty_{r=1} \beta^r Q^r_A\right) Q_A
		= Q_A + \sum^\infty_{r=1} \beta^r  Q_A^{r+1}.
	\end{equation}
	Under $\rho$-period finite dependence for choices $a \in \cA \backslash \{A\}$ and $A$ for any states $x \in \cX$, $Q_a Q_A^\rho = Q_A Q_A^\rho$ holds. As a result, equation \eqref{eq:mat-inv-1} becomes
	\begin{equation}
		\label{eq:mat-inv-1b}
		Q_a (I_X - \beta Q_A)^{-1} 
		= Q_a + \sum^{\rho - 1}_{r=1} \beta^r Q_a Q_A^r + \sum^{\infty}_{r=\rho} Q_A^{r+1}.
	\end{equation}
	Substituting equations \eqref{eq:mat-inv-2} and \eqref{eq:mat-inv-1b} into the expression in Proposition \ref{prop:nonp1} gives
	\begin{equation*}
		\frac{\partial \pi_a}{\partial \beta} = -(Q_a- Q_A)(I_X + \beta Q_A + \cdots + \beta^{\rho-1} Q_A^{\rho - 1})(I_X - \beta Q_A)^{-1} (-\log \widehat p_A),
	\end{equation*}
as desired.
\end{proof}

\begin{proof}[\textit{\textbf{Proof of Corollary \ref{prop:nonp-cor1}}}] 

Suppose that choice $A$ is a renewal action. Following the same argument as in the proof of Proposition \ref{prop:nonp2}, $I_X - \beta Q_A$ is invertible. Using equation \eqref{eq:nonp-2-1} and the fact that $Q_a Q_A = Q_A^2$ due to one-period finite dependence, the derivative can be written as
\begin{align}
	\frac{\partial \pi_a}{\partial \beta}
	& = -(Q_a - Q_A)(I_X - \beta Q_A)^{-1}(-\log p_A) \notag \\ 
	& = -(Q_a - Q_A) \left( I_X + \sum^\infty_{r=1} \beta^r Q_A^r\right) (-\log p_A) \notag \\
	& = \left[-Q_a + Q_A - \sum^\infty_{r=1} \beta^r (Q_a Q_A^r - Q_A Q_A^r )\right] (-\log p_A)  \notag \\ 
	& = (-Q_a + Q_A) (-\log p_A), \label{eq:proof-nonp3-1}
\end{align}
for any $a \in \cA \backslash \{A\}$. Note that the RHS of equation \eqref{eq:proof-nonp3-1} is independent of $\beta$. In particular, the $x$-th row of equation \eqref{eq:proof-nonp3-1} can be written as
\[
	\frac{\partial \pi_a(x)}{\partial \beta} = [-Q_a(x) + Q_A(x)]'(-\log p_A) \equiv \xi_x.
\]
Thus, $\pi_a(x)$ is increasing, decreasing, or constant in $\beta$ if $\xi_x > 0$, $\xi_x < 0$, or $\xi_x = 0$ respectively.
\end{proof}

\begin{proof}[\textit{\textbf{Proof of Corollary \ref{prop:nonp-cor2}}}] 
To begin with, the $x$-th row in equation \eqref{eq:proof-nonp3-1} can be written in summation form as
\begin{align}
	[-Q_a(x) + Q_A(x)]'(-\log p_A)
	& = \sum^{X}_{y=1} [q(y|x, A) - q(y|x, a)] [-\log p_A(y)]. \label{eq:proof-nonp3-2}
\end{align}
Since $A$ is a renewal action, the transition matrix is given by 
\[
	Q_A = \begin{pmatrix} 1 & 0 & \cdots & 0 \\ 
	1 & 0 & \cdots & 0 \\
	\vdots & \vdots & \ddots & \vdots \\
	1 & 0 & \cdots & 0 
	\end{pmatrix},
\]
i.e., all states are reset to state 1 if choice $A$ is chosen. Hence, equation \eqref{eq:proof-nonp3-2} can be further written as
\begin{align}
	[-Q_a(x) + Q_A(x)]'(-\log p_A)
	& =[-\log p_A(1)] - \sum^X_{y=1} q(y|x,a) [-\log p_A(y)].\label{eq:proof-nonp3-3}
\end{align}
Under the assumption that $0 < p_A(1) \leq p_A(y)$ for any $y \in \cX$, it follows that $-\log  p_A(1) \geq  -\log  p_A(y)$. Hence, equation \eqref{eq:proof-nonp3-3} can be written as
\begin{align}
	[-Q_a(x) + Q_A(x)]'(-\log  p_A)
	& =[-\log p_A(1)] - \sum^X_{y=1} q(y|x,a) [-\log p_A(y)] \notag \\
	& \geq [-\log p_A(1)] - \sum^X_{y=1} q(y|x,a) [-\log p_A(1)] \notag \\
	& =0, \label{eq:proof-nonp3-4} 
\end{align}
where the last line follows from the fact that each row of the transition matrix sums to 1. Since equation \eqref{eq:proof-nonp3-4} holds for any $x \in \cX$, it follows that $\pi_a$ is nondecreasing in $\beta$. \par 

Similarly, if $ p_A(1) \geq  p_A(y)$ for any $y \in \cX$, then $-\log p_A(1) \leq -\log p_A(y)$ for any $y \in \cX$. If $p_A(1) > 0$, equation \eqref{eq:proof-nonp3-3} can be written as follows
\begin{align*}
	[-Q_a(x) + Q_A(x)]'(-\log  p_A)
	& =[-\log  p_A(1)] - \sum^X_{y=1} q(y|x,a) [-\log  p_A(y)] \notag \\
	& \leq [-\log  p_A(1)]- \sum^X_{y=1} q(y|x,a) [-\log  p_A(1)] \notag \\
	& =0, 
\end{align*}
which implies that the utility is nonincreasing in $\beta$.
\end{proof}

\begin{proof}[\textit{\textbf{Proof of Proposition \ref{prop:7}}}] 
The optimization problem in \eqref{eq:min-dist-1} is analogous to least squares estimation. It follows that the solution is given by 
\begin{equation}
	\label{eq:min-dist-2}
	\widehat \theta = (\Pi'W\Pi)^{-1}(\Pi'W\widehat \pi).
\end{equation}

Note that $\Pi$ and $W$ are independent of $\beta$ by construction. Therefore, the derivative of $\widehat\theta$ with respect to $\beta$ can be obtained as follows:
\[
	\frac{\partial \widehat\theta}{\partial \beta}=
	(\Pi'W \Pi)^{-1}\Pi'W \frac{\partial \widehat \pi}{\partial \beta}.
\]

\end{proof}

\begin{proof}[\textit{\textbf{Proof of Corollary \ref{cor:6a}}}] 
By the given construction, the matrix $\Pi$ can be written as
\[
	\Pi = \begin{pmatrix} 
		\Pi_0 & 0 & \cdots & 0 \\
		0 & \Pi_1 & \cdots & 0 \\
		\vdots & \vdots & \ddots & \vdots \\ 
		0 & 0 & \cdots & \Pi_{A-1}
	\end{pmatrix}.
\]
By definition, $\widehat \pi = (\widehat \pi_0', \ldots, \widehat \pi_{A-1}')'$. Together with the assumption that $W$ is an identity matrix, the solution in equation \eqref{eq:min-dist-2} can be written as
\begin{align*}
	\widehat \theta 
	& = \begin{pmatrix}
		\Pi_0'\Pi_0 & 0 & \cdots & 0 \\
		0 & \Pi_1'\Pi_1 & \cdots & 0 \\
		\vdots & \vdots & \ddots & \vdots \\ 
		0 & 0 & \cdots & \Pi_{A-1}'\Pi_{A-1}
	\end{pmatrix}^{-1} \begin{pmatrix}
		\Pi_0'\widehat \pi_0 \\
		 \Pi_1'\widehat \pi_1  \\
		\vdots \\ 
		 \Pi_{A-1}'\widehat \pi_{A-1}
	\end{pmatrix} 	\\
	& =  \begin{pmatrix}
		(\Pi_0'\Pi_0)^{-1}\Pi_0'\widehat \pi_0 \\
		 (\Pi_1'\Pi_1)^{-1}\Pi_1'\widehat \pi_1  \\
		\vdots \\ 
		 (\Pi_{A-1}' \Pi_{A-1})^{-1} \Pi_{A-1}'\widehat \pi_{A-1}
	\end{pmatrix}.
\end{align*}	
Taking the derivative of the $a$th block above with respect to $\beta$ gives
\[
	\frac{\partial \widehat\theta_a}{\partial \beta}=
	(\Pi_a' \Pi_a)^{-1}\Pi_a' \frac{\partial \widehat \pi_a}{\partial \beta},
\]
for $a \in \cA \backslash \{A\}$.
\end{proof}

\begin{proof}[\textit{\textbf{Proof of Proposition \ref{prop:8}}}] 

To begin with, note that
\begin{align}
	\frac{\partial (\Pi'W\Pi)^{-1}}{\partial \delta}
	& = -(\Pi'W\Pi)^{-1} \frac{\partial (\Pi' W\Pi)}{\partial \delta} (\Pi'W\Pi)^{-1} \notag \\ 
	& = -(\Pi'W\Pi)^{-1} \Pi' (W' + W) \frac{\partial \Pi}{\partial\delta} (\Pi'W\Pi)^{-1}. \label{eq:prop8-1}
\end{align}
Hence, taking the derivative of equation \eqref{eq:min-dist-2} with respect to $\delta$ gives
\begin{align*}
	\frac{\partial \widehat \theta}{\partial \delta}
	& = -(\Pi'W\Pi)^{-1} \frac{\partial (\Pi' W\Pi)}{\partial \delta} (\Pi'W\Pi')^{-1}(\Pi'W\widehat \pi)
	+ (\Pi'W\Pi)^{-1} \left(\frac{\partial \Pi}{\partial \delta}\right)'W\widehat \pi \\
	& = -(\Pi'W\Pi)^{-1} \Pi' (W' + W) \frac{\partial \Pi}{\partial\delta} \widehat \theta
	+ (\Pi'W\Pi)^{-1} \left(\frac{\partial \Pi}{\partial \delta}\right)'W\widehat \pi  \\
	& = -(\Pi'W\Pi)^{-1} \left[ \Pi' (W' + W) \frac{\partial \Pi}{\partial\delta} \widehat \theta
	- \left(\frac{\partial \Pi}{\partial \delta}\right)'W\widehat \pi   \right],
\end{align*}
where the second equality uses equations \eqref{eq:min-dist-2} and \eqref{eq:prop8-1}.
\end{proof}

\section{More details on Chen and Choo (2023)} \label{app:B} 
\begin{figure}[H]
	\centering
	\caption{Actual marital surplus differences against approximation error (in percentage) for various discount factors $\beta$.}
	\label{fig:cc-supp-1}
	\includegraphics[scale=1]{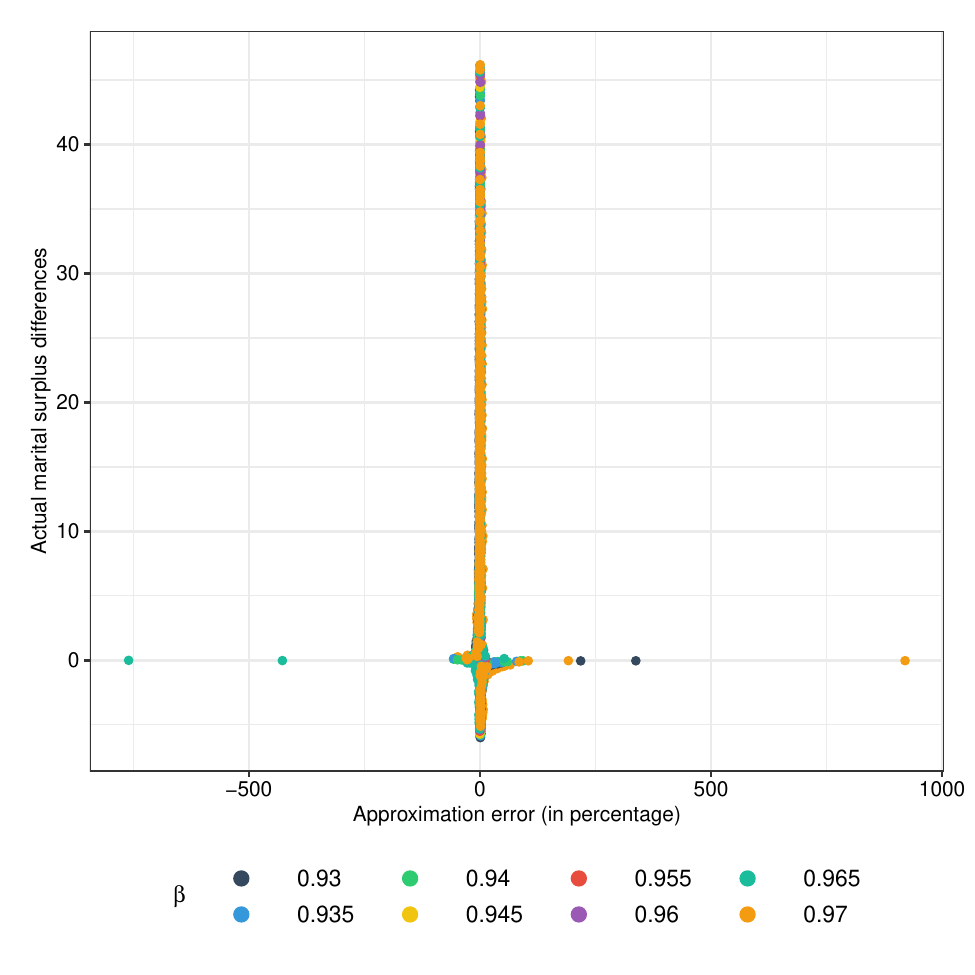}
\end{figure}

This appendix provides additional details on the approximation error of local approximation using the local sensitivity measure for the empirical application in Section \ref{sec:cc}. Figure \ref{fig:cc-supp-1} shows the scatter plot of actual marital surplus differences against the approximate error (in percentage) for different discount factors $\beta$. It can be seen that the large percentage of approximation errors are associated with actual marital surplus differences being very close to 0 for various $\beta$. \par
\begin{figure}[H]
	\centering
	\caption{Actual marital surplus differences against approximation error for various discount factors $\beta$.}
	\label{fig:cc-supp-2}
	\includegraphics[scale=1]{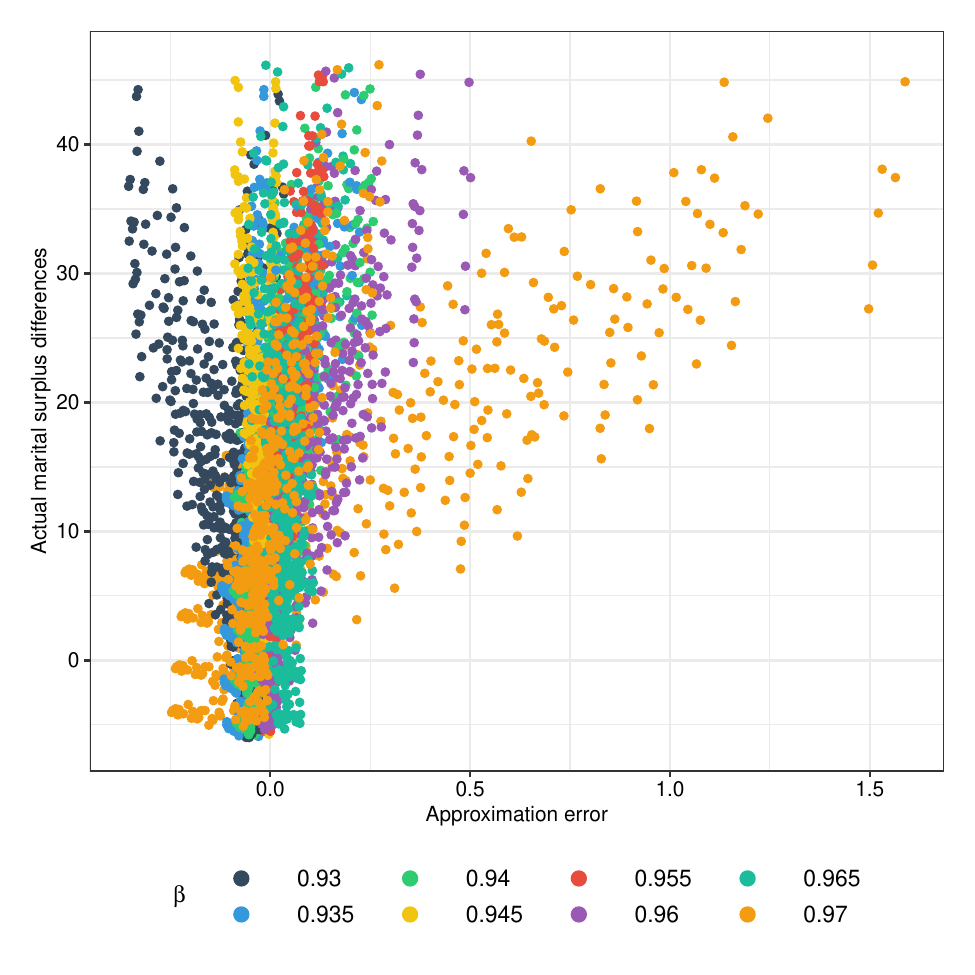}
\end{figure}

 
Figure \ref{fig:cc-supp-2} shows the scatter plot of actual marital surplus differences against the approximate error for different $\beta$. This figure shows that the approximate error remains small for various $\beta$.

\newpage
\addcontentsline{toc}{section}{References}
{\small{
	\singlespacing{
	\bibliographystyle{ecta}
	\bibliography{myref}
	}
}}

\end{document}